\documentclass[aps,pra,twocolumn,showpacs,superscriptaddress,groupedaddress]{revtex4}  
\usepackage{graphicx}  
\usepackage{dcolumn}   
\usepackage{bm}        
\usepackage{amssymb}   
\usepackage{amsmath}   
\newcommand{\bra}[1]{\ensuremath{\left\langle#1\right|}}
\newcommand{\ket}[1]{\ensuremath{\left|#1\right\rangle}}

\makeatletter
\newcommand*{\rom}[1]{\expandafter\@slowromancap\romannumeral #1@}
\makeatother

\begin{document}

\title{Non-Linear Compton Scattering in a Strong Rotating Electric Field}
\author{Erez Raicher\footnote{E-mail address: erez.raicher@mail.huji.ac.il }}
\affiliation{Racah Institute of Physics, Hebrew University, Jerusalem 91904, Israel }
\affiliation{Department of Applied Physics, Soreq Nuclear Research Center, Yavne 81800, Israel }
\author{Shalom Eliezer}
\affiliation{Department of Applied Physics, Soreq Nuclear Research Center, Yavne 81800, Israel }
\affiliation{Nuclear Fusion Institute, Polytechnic University of Madrid, Madrid, Spain }
\author{Arie Zigler}
\affiliation{Racah Institute of Physics, Hebrew University, Jerusalem 91904, Israel }

\date{\today}

\begin{abstract}
The non-linear Compton scattering rate in a rotating electric field is explicitly calculated for the first time. For this purpose,
a novel solution to the Klein-Gordon equation in the presence of a rotating electric field is applied. An analytical expression for the emission rate is obtained, as well as a simplified approximation adequate for emplementation in kinetic codes. The spectrum is numerically calculated for nowadays optical and X-ray laser parameters. The results are compared to the standard Volkov-Ritus rate for a particle in a plane wave, which is commonly assumed to be valid for a rotating electric field under certain conditions.
Subsequent deviations between the two models, both in the radiated power and the spectral shape, are demonstrated. 
First, the typical number of photons participating in the scattering process is much smaller compared to the Volkov-Ritus rate, resulting in up to an order of magnitude lower emitted power.
Furthermore, our model predicts a discrete harmonics spectrum for electrons with low asymptotic momentum compared to the field amplitude. This discrete structure is a clear imprint of the electric field frequency, as opposed to the Volkov-Ritus rate which reduces to the constant crossed field rate for the physical conditions under consideration.
Our model predictions can be tested with present-days laser facilities.

 \end{abstract}

\pacs{12.20.Ds, 52.38.-r}
\maketitle


\section{Introduction}
The interaction of electromagnetic fields with matter is one of the most fundamental problems in physics.
The conventional way to introduce the interaction with the photon into the matter equation of motion is by perturbation theory. 
For this attitude to be adequate, the interaction term should be small with respect to the other terms in the Hamiltonian.
However, if the amplitude of the electromagnetic field under consideration exceeds a certain value, a different framework must be adopted. 
In order to quantitatively characterize the transition to the strong field regime (where the standard perturbation theory fails), the non-linearity parameter is introduced
\begin{equation}
\xi \equiv \frac{ea}{m}.
\label{eq:av xi_def}
\end{equation}
Natural units are used ($\hbar = c = 1$), e and m are the electron charge and mass respectively
and $a \equiv \sqrt{-A_{\mu} A^{\mu}}$ is the amplitude of the vector potential $A_{\mu}$.
The intuitive interpretation of $\xi$, which is the reciprocal of the known Keldysh parameter \cite{keldysh, nikishov}, is the typical number of photons participating in the scattering process of the particle and the elecromagnetic wave (from now on the initials EM shall be used). As a result, if $\xi \gg 1$, i.e. in the strong-field regime, it involves many photons absorption. 
In the opposite case ($\xi < 1$), known as the perturbative regime, a non-linear process is possible but the rate $W$ decreases sharply with $n$ (the number of the participating photons), namely $W^{(n)} \propto \xi^{2n}$.
In practical units the non-linearity parameter is given by $\xi = 7.5 \sqrt{I[10^{20}W/cm^2]}/\omega[eV]$, where $I,\omega$ are the laser intensity and frequency respectively.

The failure of the standard perturbation technique in the strong field regime calls for a nonperturbative formalism. The essence of the nonperturbative attitude (also known as the strong field approximation) is that instead of treating the laser background perturbatively, we include it in the free Lagrangian \cite{Furry}. Therefore, nonperturbative calculations of QED processes in the presence of a laser field ("laser assisted") are carried out by replacing the free electron wavefunction appearing in the quantum calculation with the solution of a particle interacting with an EM plane wave traveling in vacuum (known as the Volkov wavefunction \cite{volkov, landau}).

The experimental exploration of the strong field regime became feasible due to the invention of the Chirped Pulse Amplification (CPA) technique 30 years ago \cite{CPA}. Since then, the laser intensity has increased in 8 orders of magnitude to the up-to-date record \cite{yanovsky} of $10^{22}W/cm^2$ at infra-red wavelength ($\omega =1.6eV$), corresponding to $\xi = 50$.
Several laser infrastructures with an expected intensity of $10^{24}-10^{25}W/cm^2$ are under construction worldwide, among which the 3 facilities of the ELI project \cite{ELI}. Several others are in the planning phase, such as the XCELS \cite{XCELS} in Russia, HiPER \cite{HIPER} in the UK and GEKKO EXA \cite{GEKKO} in Japan.

Concurrently, a breakthrough in free electron laser physics during the 80's made it possible to achieve intense coherent X ray light. Nowadays there are several operating XFEL facilities (e.g. LCLS in Standford \cite{LCLS}, SACLA in Japan \cite{SACLA} and FLASH in Hamburg \cite{FLASH}) and one under construction (XFEL in Hamburg \cite{XFEL}).
The maximum intensity produced in these facilities lies in the range $10^{20} - 10^{21}W/cm^2$, corresponding to $\xi \approx 2 \cdot 10^{-3}$.

The experimental availability of such intense field sources creates exciting opportunities in many research fields related to strong field physics \cite{Tajima, mendonca2}, such as attosecond spectroscopy \cite{Krausz}, relativistic nonlinear optics and relativistic high-order harmonic generation \cite{atto, bulanov}, ultrastrong laser-plasma interaction and particle acceleration \cite{esarey, macchi}, laboratory astrophysics \cite{chen}, laser-assisted QED processes \cite{DiPiazza, ritus}, Schwinger pair production \cite{schwinger} and exotic nuclear physics \cite{zamfir}.

This work is devoted to one of the most significant laser-assisted QED processes - the non-linear Compton scattering. 
Unlike the standard Compton process, where a photon scatters off an electron, the non-linear Compton describes the coherent interaction of many photons with an electron. The outcoming particles are the electron and a single energetic photon. This scattering is of particular significance for several reasons. First, it may be used to create gamma sources in the interaction of ultra-intense lasers with an energetic electrons beam. Second, it is one of the main processes responsible to the electron radiation losses in the interaction of ultra-intense lasers with plasma. In particular, the interplay between this process and the Breit-Wheeler process, involving a hard photon interacting with many laser photons to create an electron-positron pair, may result in a mechanism called "QED cascade" in the following way. The hard photon emitted during the non-linear Compton decays into an electron-positron pair through the Breit-Wheeler scattering. The newly born particles also radiate hard photons through the non-linear Compton, leading to the emergence of an avalanche.
These QED cascades attract increasing scientific attention \cite{sokolov_prl, sokolov, pukhov, nerush, nerush2, ridgers, ridgers2, brady, liang, fedotov, elkina} both for fundamental and practical causes. Practically, spontaneous cascades may drain energy from the laser pulse and thus limit the utmost attainable intensity \cite{fedotov}. From fundamental point of view the cascades are of interest as they result in a QED plasma (namely electrons, positrons and gamma photons) resembling many astrophysical scenarios \cite{liang}.

The most favorable configuration to achieve the QED cascade is a rotating electric field \cite{elkina}. It may realized in the vicinity of the antinodes of a standing wave formed by two counterpropagating laser beams. The standard kinetic modeling is consisted of a PIC code to describe the plasma motion combined with a Monte Carlo QED module to account for the strong field QED emission processes listed above. The QED rates are those obtained by the Volkov wavefunction, though the EM field configuration is different than the one used in the Volkov derivation.

The justification to this approximation was formulated in the first treatment of the non-linear Compton, carried out by Ritus, Nikishov and Narozhny \cite{nikishov, narozh} in the 60's (see also the comprehensive review \cite{ritus}). Their derivation is established on the Volkov solution and will be referred from now on as "Volkov-Ritus". The explanation of their argument requires the introduction of the 4 dimensionless quantities on which the quantum rate depends. The first is the field strength $\xi$ introduced above. The second is the normalized acceleration experienced by the particle in its rest frame. It is known as the quantum parameter and takes the form 
\begin{equation}
\chi \equiv \frac{e}{m^3} \sqrt{-(F^{\mu \nu} \Pi_{\nu})^2},
\label{eq:av chi}
\end{equation}
where $\Pi_{\nu}$ is the eigenvalue of the kinetic momentum operator $-i\partial_{\mu}-eA_{\mu}$ and the EM field strength tensor is given by
\begin{equation}
F_{\mu \nu} = {\partial}_\mu A_{\nu} - {\partial}_\mu A_{\nu}.
\end{equation}
The classical regime, i.e. the non-linear Thomson scattering, corresponds to $\chi \ll 1$. The next generation lasers are expected to enter the quantum regime, $\chi \approx 1$.
Two additional quantities are the EM field invariants
\begin{equation}
\mathcal{F} \equiv \frac{e^2F_{\mu \nu}F^{\mu \nu}}{4m^4}, \quad \mathcal{G} \equiv \frac{\epsilon_{\alpha \beta \mu \nu} e^2F^{\alpha \beta}F^{\mu \nu}}{4m^4}.
\end{equation}
The symbol $\epsilon_{\alpha \beta \mu \nu}$ stands for the Levi-Civita tensor.
Ritus and Nikishov argued that as long as the following conditions hold
\begin{equation}
\mathcal{F},\mathcal{G} \ll \chi^2, \quad \mathcal{F},\mathcal{G} \ll 1, \quad \xi \gg 1
\label{eq:av assump}
\end{equation}
the rate is well described by the Volkov-Ritus expression (coinciding under these conditions with the emission in a constant crossed field).

However, it was recently demonstrated by the authors \cite{myPaper3} that the wavefunction of a particle in a rotating electric field exhibits significant deviation from the Volkov solution even if (\ref{eq:av assump}) is satisfied. Consequently, we are motivated to explore the emission rate corresponding to our new wavefunction as compared to the familiar Volkov-Ritus rate. For the sake of simplicity, the investigation was carried out for the scalar case, neglecting the spin effects. These were shown to be of secondary importance for ultra-intense laser particle interaction \cite{elotz}.

The paper is organized as follows. In Sec. \rom{2} the strong field Lagrangian is written down and the second quantization is outlined.
Sec. \rom{3} reviews the analytical solution derived in \cite{myPaper3} for a particle in a rotating electric field. Sec. \rom{4} describes the phase space factor appearing in the non-linear Compton scattering rate. Sec. \rom{5} includes the detailed calculation of the matrix element. In Sec. \rom{7} we explicitly show that under a certain condition our formula recovers the Volkov-Ritus one. Sec. \rom{8} deals with a continuum approximation to the new rate, and Sec. \rom{6} contains the final expression for the emission spectrum. In Sec. \rom{9} the new rate is evaluated numerically and compared to the Volkov-Ritus expression for physical parameters corresponding to present-days laser facilities. Sec. \rom{10} concludes the paper.

\section{The Lagrangian Formulation}
The final goal of this work is the calculation of the non-linear Compton scattering rate in a rotating electric field. 
For this purpose, a Lagrangian formulation of the problem, including second quantization, is required. This framework, known as strong field QED, was developed for the Volkov problem long ago \cite{Furry, ritus} and was recently generalized by the authors \cite{myPaper2} for the case of
a rotating electric field.
The main results of the generalization are given below.

The Lagrangian of the scalar QED reads
\begin{multline}
\mathcal{L}_{sQED} =\frac{1}{2}\partial_{\mu}\Phi^* \partial^{\mu}\Phi  -  \frac{1}{2} m^2 \Phi^* \Phi  \\ -  \frac{1}{16 \pi}F_{\mu \nu}F^{\mu \nu}  + A_{\mu} \cdot ( f^{\mu}_1 + A^{\mu} f_2),
\label{eq:av sLag}
\end{multline}
where $\Phi$ is the scalar field operator and the center dot stands for Lorentz contraction.
The last term in the Lagrangian, representing the interaction between light and matter, is expressed using the following definitions,
\begin{equation}
f^{\mu}_1 \equiv  \frac{1}{2} ie\left(  \Phi^* \partial^{\mu} \Phi - \Phi \partial^{\mu} \Phi^*   \right) 
\label{eq:av f1_def}
\end{equation}
and
\begin{equation}
f_2 \equiv \frac{1}{2} e^2 |\Phi|^2
\label{eq:av f2_def}
\end{equation}
In standard QED, the interaction term in (\ref{eq:av sLag}) is considered as a perturbation. However, in the presence of strong field, $A_{\mu}$ acquires a vacuum expectation value and the interaction term should be redefined according to Furry \cite{Furry},
\begin{equation}
A_{\mu} = A^{cl}_{\mu} + A^{Q}_{\mu}, \quad A^{cl}_{\mu}   \equiv \bra{\Omega}  A_{\mu}  \ket{\Omega},
\label{eq:av A_decomp}
\end{equation}
where $\ket{\Omega}$ stands for the vacuum state and will be defined below.
We substitute (\ref{eq:av A_decomp}) into (\ref{eq:av sLag}) and group all terms involving both $A^Q_{\mu}$ and $f^{\mu}_1,f_2$.
\begin{equation}
\mathcal{L}_{int} = 2 f_2 \left(  A^{cl}  \cdot A^Q  \right)  +  A^Q \cdot f_1  +  \left( {A^Q}\right)^2   f_2. 
\label{eq:av LintS1}
\end{equation}
The remaining terms are included in the free part of the Lagrangian 
\begin{equation}
\mathcal{L}_{free} =\frac{1}{2}\partial_{\mu}\Phi^* \partial^{\mu}\Phi  -  \frac{1}{2} m^2 \Phi^* \Phi  \\ -  \frac{1}{16 \pi}F_{\mu \nu}F^{\mu \nu}   + \mathcal{L}_{Furry},
\label{eq:av free4}
\end{equation}
where
\begin{equation}
\mathcal{L}_{Furry} = A^{cl} \cdot f_1 + f_2 {A^{cl}}^2.
\end{equation}
Finally, the full Lagrangian is the sum $\mathcal{L} = \mathcal{L}_{free} + \mathcal{L}_{int}$.

The free equations of motion corresponding to (\ref{eq:av free4}) are
\begin{equation}
\left [- {\partial}^2 + e^2{A^{cl}}^2 - 2ie A^{cl} \cdot \partial -m^2 \right] \Phi =  0,
\label{eq:av KG3}
\end{equation}
\begin{equation}
\partial^2 A^Q_{\mu} =0,
\label{eq:av Max5}
\end{equation}
\begin{equation}
\partial^2 A^{cl}_{\mu} =0.
\label{eq:av Max5}
\end{equation}

The solution of the system, addressed in the next section, enables us to proceed with the second quantization procedure 
\begin{equation}
\Phi =  \int{ \frac{d^3p}{(2 \pi)^{\frac{3}{2}} \sqrt{2q_0}} \left(  c_q \phi_A (q) + h.c \right) }
\label{eq:av phi_series1}
\end{equation}
\begin{equation}
A^Q_{\mu} =  \int{ \frac{d^3k'}{(2 \pi)^{\frac{3}{2}} \sqrt{2k'_0}} \left[ \epsilon'_{\mu} a_{k'} e^{-i k' \cdot x}+ h.c  \right]} 
\label{eq:av A_quant}
\end{equation}
where $\phi_A (p),\epsilon'_{\mu} e^{-i k' \cdot x}$ are the one particle solution of (\ref{eq:av KG3}) and (\ref{eq:av Max5}) respectively.
We introduced the creation and annihilation operators, obeying the following cummutations relations
\begin{equation}
[a^{\dagger}_{k'},a_{k''}] = (2 \pi)^3 \delta^3(k'' - k')
\end{equation}
\begin{equation}
[c^{\dagger}_q,c_{q'} ] = (2 \pi)^3 \delta^3(q - q')
\end{equation}

Now let us specify the vacuum state $\ket{\Omega}$ and the corresponding classical field $A^{cl}_{\mu}$.
As was mentioned in the introduction, a rotating electric field can be realized at the antinodes of a standing wave formed by colliding circularly polarized laser beams. 
The vector potential corresponding to this configuration is
\begin{multline}
A^{cl,\mu} = \frac{1}{2} a^{\mu}_1 \left[ \cos (k_1 \cdot x) + \cos (k_2 \cdot x) \right] + \\ \frac{1}{2} a^{\mu}_2 \left[  \sin (k_1 \cdot x) +\sin (k_2 \cdot x)  \right]
\label{eq:av A_counter}
\end{multline}
where the wave vectors are $k^{\mu}_1 = (\omega,\textbf{k})$ and $k^{\mu}_2 = (\omega,-\textbf{k})$ and satisfy the vacuum dispersion relation $k^2_1 = k^2_2 = 0$. The polarization vectors are given by
\begin{equation}
a^{\mu}_1 = a \hat{e}^{\mu}_x, \quad a^{\mu}_2 = a \hat{e}^{\mu}_y
\label{eq:av a12_def}
\end{equation}
and the unit vectors read
\begin{equation}
\hat{e}^{lab}_1 = (0,1,0,0), \quad \hat{e}^{lab}_2 = (0,0,1,0).
\end{equation}
where the superscript "lab" attached to a 4-vector denotes that it is evaluated in the laboratory frame of reference. Notice that $A^{cl}$ given by (\ref{eq:av A_counter}) satisfies, as it should, the relevant equation of motion (\ref{eq:av Max5}).
The ground state corresponding to this field configuration is
\begin{equation}
\ket{\Omega}= \ket{0}\ket{\alpha,k_1; \alpha, k_2}
\end{equation}
where \ket{0} is the scalar part of the wave function and $\ket{\alpha,k_1; \alpha, k_2}$ stands for two coherent states with 4-momenta $k_1,k_2$ respectively, representing the counterpropagating laser beams.

\section{Analytical solution}
In the previous section, the equation of motion describing the dynamics of the scalar field operator was derived (\ref{eq:av KG3}). Substituting the ansatz (\ref{eq:av phi_series1}) for $\Phi$, one may obtain the equation satisfied by the wavefunction $\phi_A (p)$ 
\begin{equation}
\left[-\partial^2-2ie(A^{cl}\cdot\partial)+e^2{A^{cl}}^2-m^2 \right]\phi_A = 0,
\label{eq:av KG_i}
\end{equation}
which is the familiar Klein-Gordon equation in the presence of classical EM field.
Using simple trigonometric manipulations, the laser field (\ref{eq:av A_counter}) takes the form
\begin{equation}
A^{cl,\mu}  =  \cos( \textbf{k}  \cdot \textbf{x}) \left[ a^{\mu}_1 \cos (\omega t) + a^{\mu}_2 \sin (\omega t)  \right]
\label{eq:av A_cl1}
\end{equation}
in the vicinity of the antinode, i.e. $\textbf{k}  \cdot \textbf{x} = 0$, the cosine equals to unity up to a second order correction, i.e. $\cos( \textbf{k}  \cdot \textbf{x}) \approx 1$.
In order to cast Eq. (\ref{eq:av A_cl1}) into a Lorentz-invariant form, namely 
\begin{equation}
A^{cl,\mu} = a^{\mu}_1 \cos (k \cdot x) + a^{\mu}_2 \sin (k \cdot x),
\label{eq:av A_cl2}
\end{equation}
a new wave vector $k_{\mu}$ is introduced. In the laboratory frame it reads $k^{lab} = (\omega,0,0,0)$. Consequently, it obeys a massive-like dispersion relation
\begin{equation}
k^2  = m_{ph}^2,
\label{eq:av dispersion}
\end{equation} 
where $m_{ph}$ is the effective mass of the rotating electric field photons. As can be inferred from the definition of $k$, the photon effective mass equals to the laser frequency in the laboratory frame, $m_{ph} =  \omega$.

The approximated wavefunction $\phi_A(q)$ solving (\ref{eq:av KG_i}) was recently published by the authors \cite{myPaper3},
\begin{multline}
\phi_A(q)=    \exp \bigl[ -iq \cdot x -i\frac{e (a_1 \cdot q)}{ k \cdot q} \sin (k \cdot x) \\+ i\frac{e (a_2 \cdot q)}{ k \cdot q} \cos (k \cdot x) \bigr].
\label{eq:av psi_ana}
\end{multline}
The quasi momentum is defined as 
\begin{equation}
q_{\mu} \equiv p_{\mu} - \nu k_{\mu},
\label{eq:av q_def}
\end{equation}
where $p_{\mu}$ is the asymptotic momentum (the momentum in the absence of the EM wave) and $\nu$ is the characteristic exponential given by
\begin{equation}
\nu = \frac{ k \cdot p  }{ {m_{ph}}^2} \left ( 1 -  \sqrt{1 + \left (  \frac{ ea {m_{ph}} }{ k \cdot p } \right )^2 }  \right ).
\label{eq:av nu_kg}
\end{equation}
In the limit $m_{ph} \rightarrow 0$ the characteristic exponential reduces to $\nu^V = -(ea)^2 / [2(k \cdot p)]$ and the Volkov solution is recovered.

The underlying assumption behind this approximate solution is
\begin{equation}
\delta \equiv \frac{eam_{ph}^2 |p \cdot (\hat{e}_x + i \hat{e}_y)|}{(k \cdot q)^2} \ll 1.
\label{eq:av delta_def}
\end{equation}
Equivalently, in the laboratory frame, it takes the form
\begin{equation}
\delta \equiv \frac{ea p^{lab}_{\bot}}{(ea)^2 + \left( p^{lab}_0 \right)^2} \ll 1,
\label{eq:av delta_def2}
\end{equation}
where $p_{\bot} \equiv  \sqrt{p_x^2+p_y^2}$ is the perpendicular asymptotic momentum.
The physical meaning of this condition is that our approximation is valid unless the perpendicular asymptotic momentum is comparable to both the field amplitude and the asymptotic energy (i.e. $p^{lab}_{\bot} \approx p^{lab}_0 $ as well as $p^{lab}_{\bot} \approx ea$).

Let us calculate three significant quantites associated with the wavefunction. The first is the dressed electron effective mass, frequently encountered in the next sections. Using the definition $m_* \equiv \sqrt{q^2} $ and (\ref{eq:av q_def}), one obtains
\begin{equation}
m_* = m\sqrt{1+\xi^2},
\label{eq:av m_star}
\end{equation}
which is identical to the one corresponding to the Volkov wavefunction \cite{ritus}.
The second quantity is the eigenvalue of the kinetic momentum operator, satisfying 
\begin{equation}
\left(-i \partial_{\mu}-eA_{\mu} \right)\phi_A = \Pi_{\mu} \phi_A.
\end{equation}
Evaluating the left side of the above equation, one finds
\begin{equation}
\Pi_{\mu} = p_{\mu} - eA_{\mu}-\nu k_{\mu},
\label{eq:av Pi_f}
\end{equation}
which is nothing but the classical momentum of a particle in a rotating electric field.
The third quantity to be calculated is the quantum parameter $\chi$, obtained by substituting (\ref{eq:av Pi_f}) into (\ref{eq:av chi}),
\begin{equation}
\chi =\frac{ea(k \cdot q)}{m^3} \sqrt{ 1 - \frac{m_{ph}^2 ({A^{cl}}' \cdot p)^2}{(ea)^2 (k \cdot q)^2} }
\label{eq:av chi_gen1}
\end{equation}
where ${A^{cl}}'$ denotes the first derivative of ${A^{cl}}$ with respect to $(k \cdot x)$.
Notice that $\chi$ is time-dependent (through ${A^{cl}}'$) unless the asymptotic momentum satisfies ${A^{cl}}' \cdot p =0$.

\section{the scattering phase space }
In the following, the non-linear Compton scattering rate corresponding to our new solution is obtained.
According to the standard formulation, we start with the transition amplitude between the initial and final state, and relate it to the interaction Lagrangian. Afterwards, the rate is written as a matrix element integrated over the available phase space of the outcoming particles. This section mainly deals with the particulars of the phase space integration, while the matrix element calculation is addressed in the next one. 

The transition amplitude $iT$ between the initial and final states reads
\begin{equation}
i T = \bra{q',k'}(\mathcal{S}-1)\ket{q},
\end{equation}
where the canonical normalization is used for the one particle states, i.e. $\ket{q} \equiv \sqrt{2q_0} c_q^{\dagger}(q) \ket{\Omega} $.
The scattering matrix in the interaction picture is given by
\begin{equation}
\mathcal{S}= \mathcal{T}e^{i\int {d^4x \mathcal{L}_{int}}},
\end{equation}
where $\mathcal{T}$ is the time ordering operator and $\mathcal{L}_{int}$ was given in (\ref{eq:av LintS1}). The non-linear Compton scattering originates in the first order term in the Taylor expansion of $\mathcal{S}$. Hence,
\begin{equation}
i T = \bra{q',k'}\int{d^4x}\mathcal{L}_{int}\ket{q}.
\label{eq:av T_mat}
\end{equation}
Omitting the term in (\ref{eq:av LintS1}) involoving $\left( A^Q_{\mu} \right)^2$ (since it does not contribute to this scattering) and writing explicitly $f^{\mu}_1,f_2$ we obtain
\begin{equation}
\mathcal{L}_{int} =  eA^{Q,\mu}\left[  \Phi^* (i \partial _{\mu} + eA^{cl}_{\mu}) \Phi  -  \Phi (i \partial _{\mu} - eA^{cl}_{\mu}) \Phi^* \right].
\label{eq:av Lag2}
\end{equation}
Substituting (\ref{eq:av phi_series1},\ref{eq:av A_quant},\ref{eq:av Lag2}) into (\ref{eq:av T_mat}) the transition amplitude takes the form
\begin{multline}
iT = ie \int{dx^4 (\epsilon'_{\mu})^*    e^{ik' \cdot x} \bigl[   {\phi}^*_{A}(q') (i\partial^{\mu} + eA^{cl}_{\mu})  \phi_A(q)} - \\ {\phi}_{A}(q)(i\partial^{\mu}-eA^{cl}_{\mu})  \phi^*_A(q') \bigr].
\label{eq:av trans_amp1}
\end{multline}
As shown later on, the integration over $d^4x$ results in an infinite sum of energy-momentum conservation delta functions 
\begin{equation}
iT = i \sum_s{}\mathcal{M}_s (2 \pi)^4 \delta^4(q + sk - k' - q'),
\end{equation}
where $\mathcal{M}_s$ is the matrix element and $s$ is the number of laser photons absorbed in the process.
Let us consider a process with a given s. The corresponding energy-momentum conservation reads
\begin{equation}
sk+q = q'+k'.
\label{eq:av delta_cons}
\end{equation}
It is favorable to hold our discussion in the center of mass (c.m.s) frame.
Let us write down the incoming 4-momenta explicitly
\begin{equation}
sk = (\sqrt{p_{in}^2+s^2m_{ph}^2},0,0,p_{in})
\label{eq:av k_init}
\end{equation}
\begin{equation}
q = (\sqrt{p_{in}^2+m_*^2}, 0,0, -p_{in} )
\label{eq:av q_init}
\end{equation}
where $p_{in}$ is the momentum of each of the incoming particles in the c.m.s frame. The 4-momenta of the outcoming particles are
\begin{equation}
q' = p_{out} \left( \sqrt{1+\left[ \frac{m_*}{p_{out}} \right]^2},  \sin \theta \cos \varphi, \sin \theta \sin \varphi,  \cos \theta \right)
\label{eq:av q_tag1}
\end{equation}
\begin{equation}
k' = p_{out} \left( 1, - \sin \theta \cos \varphi, - \sin \theta \sin \varphi, - \cos \theta \right)
\label{eq:av k_tag}
\end{equation}
where $\theta, \varphi$ are the scattering angles and $p_{out}$ is the momentum of the outcoming particles in the c.m.s frame.
The c.m.s energy is given by
\begin{equation}
E_s^2 = (sk+q)^2 = s^2 m_{ph}^2 + m_*^2 + 2sk \cdot q.
\label{eq:av Es2_def}
\end{equation}
The initial c.m.s. momentum is related to $E_s$ by
\begin{equation}
E_s = sk_0 +q_0 =  \sqrt{s^2m_{ph}^2+p_{in}^2}+\sqrt{m_*^2+p_{in}^2}.
\label{eq:av p_in0}
\end{equation}
The solution of equation (\ref{eq:av p_in0}) yields 
\begin{equation}
p_{in} = \frac{s(k \cdot q)}{E_s} \sqrt{1 - \left( \frac{m_* m_{ph}}{k \cdot q} \right)^2 }.
\label{eq:av p_in}
\end{equation}
The final c.m.s. momentum is related to $E_s$ by
\begin{equation}
E_s = p_{out} + \sqrt{  m_*^2 + p_{out}^2 }.
\label{eq:av p_out0}
\end{equation}
Hence we have
\begin{equation}
p_{out} = \frac{E_s^2 - m_*^2}{2E_s}=\frac{s^2 m_{ph}^2 + 2s k \cdot q}{2E_s}.
\label{eq:av p_out}
\end{equation}
For vanishing $m_{ph}, $ Eqs. (\ref{eq:av p_in0}, \ref{eq:av p_out0}) take exactly the same form. Therefore, Eqs. (\ref{eq:av p_in}, \ref{eq:av p_out}) become identical ($p_{in} = p_{out}$) as expected.

In order to obtain the total emission rate for a hard photon by the electron, the phase space integration should be performed. The rate is related to the transition amplitude by \cite{landau}
\begin{equation}
W = \frac{1}{2q_0} \int{\frac{d^3q'}{(2 \pi)^3 2q'_0}\frac{d^3k'}{(2 \pi)^3 2k'_0} |T|^2}.
\label{eq:av rate1}
\end{equation}
Notice that the integration result is Lorentz-invariant and the only frame-dependent term is the coefficient $1/(2q_0)$ multiplying the integral.
The delta function appearing in the expression for $|T|^2$ contains 4 constraints on the possible final states.
Therefore the summation over the phase space reduces into a two dimensional integral \cite{peskin}
\begin{equation}
\delta^4(sk + q -q' -k')\frac{d^3q'd^3k'}{q_0'k_0'} \rightarrow \frac{p_{out}d(\cos \theta) d\varphi}{E_s}.
\label{eq:av phase_int}
\end{equation}

Since each $s$ has a unique c.m.s. frame, different $s$ corresponds to different $\theta$. As a result, we would like to replace $\cos \theta$ in (\ref{eq:av phase_int}) by more a convenient variable. For this purpose, a new Lorentz-invariant parameter is introduced
\begin{equation}
u \equiv \frac{k \cdot k'}{k \cdot q'}.
\label{eq:av u_def1}
\end{equation}
This variable is also an indicator to the classical / quantum nature of the scattering \cite{ritus}. Since $u \ll 1$ implies a negligible momentum of the outcoming photon, it corresponds to the classical limit. On the other hand, for $u \approx 1$ the quantum mechanics dominates the process.
One may show that $\theta $ is related to $u$ by (see Appendix A)
\begin{equation}
\cos \theta = \eta_s \left( \kappa_s - \frac{\kappa_s+1}{1+u} \right)
\label{eq:av cos_u}
\end{equation}
with 
\begin{equation}
\kappa_s \equiv \frac{E_s^2 + m_*^2}{E_s^2 - m_*^2}
\label{eq:av kappa_def}
\end{equation}
and
\begin{equation}
 \eta_s \equiv \frac{sk_0}{p_{in}}.
\label{eq:av alpha_s1}
\end{equation}
Notice that in the Volkov limit (i.e. $m_{ph} \rightarrow 0$) we have $\eta_s =1$ by definition, due to the vacuum dispersion of the laser photons. In order to obtain $\eta_s$ in terms of the initial quantities, Eq. (\ref{eq:av p_in}) as well as the relation $k_0 = \sqrt{ m_{ph}^2 + (p_{in} / s)^2}$ are employed.
\begin{equation}
\eta_s =\sqrt{ \frac{(k \cdot q)^2 + 2s (k \cdot q)m_{ph}^2 + s^2 m_{ph}^4}{(k \cdot q)^2 - (m_{ph} m_*)^2} }.
\label{eq:av alpha_s2}
\end{equation}
In the lab frame, $k \cdot q = m_{ph} q^{lab}_0$. As a result, Eq. (\ref{eq:av alpha_s2}) simplifies to
\begin{equation}
\eta_s = \frac{q^{lab}_0+s m_{ph}}{\sqrt{\left( q^{lab}_0 \right)^2 - m_{*}^2}}.
\label{eq:av alpha_fin}
\end{equation}
The limiting values of $u$, corresponding to $\cos \theta = \pm 1$, are
\begin{equation}
u_{s,min} = \frac{\eta_s -1 }{ \eta_s \kappa_s + 1}, \quad u_{s,max} = \frac{\eta_s +1 }{ \eta_s \kappa_s - 1}.
\end{equation}
For vanishing $m_{ph}$ the well known Volkv-Ritus expression is recovered, namely $u^V_{s,min} = 0$ and
\begin{equation}
u^V_{s,max} = \left ( \frac{E_s}{m_*} \right )^2 - 1.
\label{eq:av u_s_v}
\end{equation}
Finally, the phase space factor (\ref{eq:av phase_int}) may be written as
\begin{equation}
\delta^4(sk + q -q' -k')\frac{d^3q'd^3k'}{q_0'k_0'} \rightarrow \frac{\eta_s dud\varphi}{(1+u)^2 }.
\label{eq:av a2_delta_ident}
\end{equation}
Substituting (\ref{eq:av a2_delta_ident}) into (\ref{eq:av rate1}) and summing over the polarization of the outcoming photon, the rate is obtained
\begin{equation}
\frac{dW}{du d \varphi} = \frac{1}{32 \pi^2 q_0} \sum_{s} \frac{\eta_s }{(1+u)^2 } \frac{1}{2} \sum_{\epsilon'}|\mathcal{M}_{s,\epsilon'}|^2.
\label{eq:av rate2}
\end{equation}
The total rate is obtained by integrating (\ref{eq:av rate2}) with respect to $u,\varphi$ in the range $0 < \varphi < 2 \pi$ and $u_{min} < u < u_{max}$. 
In order to return to $c.g.s$ unit system the simple transformations $m \rightarrow mc^2$, $m_{ph} \rightarrow m_{ph}c^2$ and $q \rightarrow \hbar q$ are carried out.

\section{Matrix element calculation}
In the above section, a relation between the transition amplitude and the particles wavefunction was established (\ref{eq:av trans_amp1}). In the following, it is further evaluated and expressed in terms of the initial quantites. 
Substituting the analytical wavefucntions (\ref{eq:av psi_ana}) into (\ref{eq:av trans_amp1}) we arrive at
\begin{multline}
i T=  ie \int  {d^4x}  (\epsilon'^{\mu})^*  e^{i Q}  \bigl[ q_{\mu} + q_{\mu}' \\ +  \left( \frac{e(q \cdot A^{cl})}{(k \cdot q)} + \frac{e(q' \cdot A^{cl})}{(k \cdot q') }   \right) k_{\mu} -2eA^{cl}_{\mu} \bigr],
\label{eq:av trans_amp2}
\end{multline}
where the exponent argument is
\begin{multline}
Q \equiv \left( q' - q + k' \right) \cdot x + \alpha_1 \sin (k \cdot x) - \alpha_2 \cos (k \cdot x)
\end{multline}
and the following quantities are introduced
\begin{equation}
\alpha_i \equiv \frac{e (a_i \cdot q)}{(k \cdot q)}-\frac{e(a_i \cdot q')}{ (k \cdot q')}, \quad i=1,2.
\label{eq:av alpha_def}
\end{equation}
It proves useful to rewrite the following expression, appearing in the exponent argument
\begin{equation}
\alpha_1 \sin (k \cdot x) - \alpha_2 \cos (k \cdot x) = z \sin \left[ (k \cdot x) - \phi_0 \right]
\end{equation}
with the definitions
\begin{equation}
z \equiv \sqrt{\alpha_1^2 +\alpha_2^2}
\label{eq:av z_def}
\end{equation}
and
\begin{equation}
\phi_0 \equiv \tan^{-1} \left(\alpha_1 / \alpha_2 \right).
\end{equation}
As shown below, the phase $\phi_0$ does not appear in the final expression and therefore bears no physical meaning.

In order to carry out the integration, we take advantage of the following identity 
\begin{equation}
(1,\cos \phi, \sin \phi)e^{iz \sin (\phi - \phi_0)} = \sum_s {(B,B_1,B_2)} e^{is \phi}	
\end{equation}
where
\begin{equation}
B \equiv J_s(z)e^{is \phi_0},
\label{eq:av B0_def}
\end{equation}
\begin{equation}
B_1 \equiv \left( \frac{s}{z}J_s(z) \cos \phi_0 + i J_s'(z) \sin \phi_0 \right) e^{is \phi_0},
\label{eq:av B1_def}
\end{equation}
\begin{equation}
B_2 \equiv \left( \frac{s}{z}J_s(z) \sin \phi_0 - i J_s'(z) \cos \phi_0 \right) e^{is \phi_0},
\label{eq:av B2_def}
\end{equation}
and $J_s(z)$ is the Bessel function. 
In terms of $B,B_1,B_2$, the integration of (\ref{eq:av trans_amp2}) yields
\begin{equation}
iT = i \sum_s{}\mathcal{M}_s (2 \pi)^4 \delta^4(q + sk - k' - q')
\end{equation}
where the matrix element takes the form
\begin{multline}
i \mathcal{M}_s =ie \epsilon'_{\mu} \bigl[( q^{\mu}+q'^{\mu} ) B  -2eB_1 a^{\mu}_1-2eB_2a^{\mu}_2 \\ +e(a_1+a_2) \cdot \left(\frac{q}{k \cdot q} +  \frac{q'}{k \cdot q'} \right) k^{\mu} \bigr]
\label{eq:av M_th}
\end{multline}
and the expression for $A_{\mu}^{cl}$ (\ref{eq:av A_cl2}) was used. The next step is to sum over the emitted photon polarization. For this purpose, we introduce the quantity $\mathcal{M}_{\mu}$, defined by
\begin{equation}
 \mathcal{M} \equiv  \mathcal{M}_{\mu} \epsilon'^{\mu}.
\label{eq:av M_mu}
\end{equation}
Due to the dispersion of the emitted photon, namely $k'^2 = 0$, one can apply the Ward identity \cite{peskin}, and therefore the summation over the outcoming photon polarization is simplified to
\begin{equation}
\sum_{\epsilon'} |\mathcal{M}|^2 = -g^{\mu \nu} \mathcal{M}_{\mu} \mathcal{M}^*_{\nu} 
\label{eq:av polar_sum}
\end{equation}
where $g^{\mu \nu} \equiv diag(1,-1,-1,-1)$ is the Minkowski metric.
Substituting (\ref{eq:av M_th}, \ref{eq:av M_mu}) into (\ref{eq:av polar_sum}) the final expression is obtained (see Appendix B for details)
\begin{multline}
\frac{1}{2}\sum_{\epsilon'} |\mathcal{M}_{s,\epsilon'}|^2 = -2e^2 J_s^2 (z) \left( 2m_*^2 + \frac{u-3}{2(u+1)} s^2 m_{ph}^2 \right) \\ + 4 e^4a^2 \left( \frac{s^2}{z^2} J_s^2(z)+ {J'}_s^2(z) \right).
\label{eq:av fin_mat}
\end{multline}

At the moment, the variable $z$ (necessery to calculate the matrix element), depends upon the following unknown quantites
\begin{equation}
(k \cdot q'), (q' \cdot a_1), (q' \cdot a_2).
\label{eq:av unknown}
\end{equation}
Let us express (\ref{eq:av unknown}) in terms of the following Lorentz-invariants
\begin{equation}
s, (k \cdot q), (q \cdot a_1), (q \cdot a_2), u, \varphi.
\label{eq:av known}
\end{equation}
We start with $k \cdot q'$. Using (\ref{eq:av u1}) from Appendix A we simply get
\begin{equation}
k \cdot q' = \frac{s m^2_{ph} + k \cdot q}{u+1}.
\label{eq:av kdq_tag}
\end{equation}

As to $(q' \cdot a_1), (q' \cdot a_2)$, a parametrization of the vector potential in the c.m.s. frame is required. We assume, without loss of generality, that the momentum of the incoming particle lies (in the lab frame) in the $x-z$ plane. Hence, in the c.m.s frame $a_2$ remains unchanged
\begin{equation}
a_2 = a (0,0,1,0).
\label{eq:av a2_gen1}
\end{equation}
and by definition $a_2 \cdot q = 0$.
In Appendix C, an expression for $a_1$ in the c.m.s. frame is derived
\begin{equation}
a_1 = a \left(R_0,\sqrt{1 - R_0^2 \left( \eta_s^2 - 1 \right)^2},0,R_0 \eta_s \right),
\label{eq:av a1_gen3}
\end{equation}
where
\begin{equation}
R_0  = \frac{q \cdot a_1} {aE_s}.
\end{equation}
Employing (\ref{eq:av a2_gen1},\ref{eq:av a1_gen3},\ref{eq:av q_tag1}), the evaluation of $(q' \cdot a_1), (q' \cdot a_2)$ is straightforward
\begin{equation}
a_2 \cdot q' = -a p_{out} \sin \theta \sin \varphi
\end{equation}
and
\begin{multline}
a_1 \cdot q' = a p_{out} \bigl[ R_0 \sqrt{1 + \left(\frac{m_*}{p_{out}} \right)} +R_0 \eta_s \cos \theta \\ -\sin \theta \cos \varphi \sqrt{1 - R_0^2 \left(\eta_s^2 - 1 \right)} \bigr].
\end{multline}
Plugging the above equations into (\ref{eq:av alpha_def}) we obtain $\alpha_1,\alpha_2$
\begin{equation}
\alpha_2 = \frac{e ap_{out} \sin \theta \sin \varphi(1+u)}{k \cdot q +s m_{ph}^2}
\label{eq:av alpha_2}
\end{equation}
and
\begin{multline}
\alpha_1 = \frac{e(a_1 \cdot q)}{k \cdot q}-\frac{eap_{out}(u+1)}{k \cdot q +s m_{ph}^2}   \bigl[ R_0 \sqrt{1 + \left(\frac{m_*}{p_{out}} \right)} \\+R_0 \eta_s \cos \theta  -\sin \theta \cos \varphi \sqrt{1 - R_0^2 \left(\eta_s^2 - 1 \right)} \bigr]
\label{eq:av alpha_1}.
\end{multline}
To sum up, the final expression for the matrix element is (\ref{eq:av fin_mat}) where $z$ is calculated using (\ref{eq:av z_def},\ref{eq:av alpha_2},\ref{eq:av alpha_1}). 

In the above derivation, $z$ depends on both $u, \varphi$, as opposed to the Volkov-Ritus case (where it depends upon $u$ only).
Notice, however, that under the condition $q \cdot a_1  = 0$ we have $R_0 = 0$ and the dependence on $\varphi$ vanishes. The expression for $z$ simplifies to
\begin{equation}
z = \frac{ea p_{out}}{sm_{ph}^2 + k \cdot q} \sin \theta (1+u).
\label{eq:av simple_z}
\end{equation}
Employing (\ref{eq:av cos_u}), the angle $\theta $ may be written in terms of $u$:
\begin{equation}
 \sin \theta (1+u) = \sqrt{\bar{A}_s u^2+\bar{B}_s u+\bar{C}_s}
\label{eq:av sin_u}
\end{equation}
with
\begin{equation}
\bar{A}_s \equiv 1-\kappa_s^2 \eta_s^2, \quad \bar{B}_s \equiv 2 (\kappa_s \eta_s^2 + 1), \quad \bar{C}_s \equiv 1 - \eta_s^2.
\label{eq:av u_pol}
\end{equation}
By equating the derivative of (\ref{eq:av sin_u}) to zero, one can readily obtain the $u$ value corresponding to the maximum of this function
\begin{equation}
u^m_{s} = \frac{\eta_s^2 \kappa_s+1}{\eta_s^4 \kappa_s^2-1}.
\label{eq:av u_m}
\end{equation}
By definition, the contribution of a given harmonic $s$ is centered around $u=u^m_s$. 
The substitution of (\ref{eq:av u_m}) into (\ref{eq:av sin_u}) yields the maximal value of this function
\begin{equation}
\bar{D}_s \equiv \sin \theta (1+u) |_{max} = \frac{\eta_s \left(\eta_s \kappa_s +1 \right)}{\sqrt{\eta^4_s \kappa^2_s -1}}.
\label{eq:av D_s}
\end{equation}

\section{the volkov-ritus limit}
The matrix element calculation being completed, a benchmark with an established result is valueable. 
For vanishing $m_{ph}$, the quantum problem is identical to the one solved by Volkov. Consequently, the results obtained in the previous section should recover the familiar Volkov - Ritus formulas \cite{ritus}. In the following, this limit is explicitly worked out.

We start with expression for the matrix element derived in the previous section (\ref{eq:av fin_mat}).
For $m_{ph}=0$, the second term vanishes and the familiar Volkov-Ritus expression for scalars \cite{ritus} is reproduced
\begin{multline}
\frac{1}{2}\sum_{\epsilon'} |\mathcal{M}_{s,\epsilon'}|_V^2 = -4e^2 J_s^2 (z)  m_*^2   \\ + 4 e^4a^2 \left( \frac{s^2}{z^2} J_s^2(z)+ {J'}_s^2(z) \right).  
\label{eq:av fin_Vol}
\end{multline}
One can observe that the Volkov-Ritus matrix element (\ref{eq:av fin_Vol}) is not very different than the obtained earlier (\ref{eq:av fin_mat}).
The major difference, however, lies in the dependence of $z$ on the particle incoming momentum, as shown below.

Let us find $z$ in the Volkov-Ritus limit. For a vanishing $m_{ph}$ we have $\eta_s=1$, as can be inspected from (\ref{eq:av alpha_s1}). Therefore, the formula (\ref{eq:av alpha_1}) for $\alpha_1$ simplifies to
\begin{multline}
\alpha_1 = \frac{e ap_{out} \sin \theta \cos\varphi}{k \cdot q} + \frac{e(a_1 \cdot q)}{k\cdot q} \bigl[ 1 -\\ \frac{(u+1)p_{out}}{E_s} \sqrt{1 + \left(\frac{m_*}{p_{out}} \right)^2} + \frac{(u+1)p_{out} \cos \theta}{E_s} \bigr].
\label{eq:av alpha_1_V}
\end{multline}
Using (\ref{eq:av cos_u_pre}) from Appendix A, the last term is rewritten
\begin{equation}
\frac{(u+1)p_{out} \cos \theta}{E_s} = \frac{(u+1)}{E_s} \left(  q_0'  -  \frac{E_s}{1+u}  \right).
\label{eq:av u_cos}
\end{equation}
Plugging (\ref{eq:av u_cos}) into (\ref{eq:av alpha_1_V}), one can observe that the last bracket is identically zero. Thus, $z$ may be cast in the form
\begin{equation}
z = \frac{ea p_{out}}{ k \cdot q} \sin \theta (1+u). 
\label{eq:av z_mid}
\end{equation}
 From $\eta_s=1$ we deduce that $\bar{C}_s = 0$ and therefore (\ref{eq:av sin_u}) simplifies to
\begin{equation}
\sin \theta (1+u) = \frac{2E_s m_*}{E_s^2 - m_*^2}\sqrt{u \left( u^V_{s,max} - u \right)}
\label{eq:av sin_v}
\end{equation}
where $u^V_{s,max}$ was defined in (\ref{eq:av u_s_v}). The Substitution of (\ref{eq:av p_out}, \ref{eq:av sin_v}) into (\ref{eq:av z_mid}) yields
\begin{equation}
z = \frac{ea m_*}{k \cdot q} \sqrt{u \left( u^V_{s,max} - u \right)}.
\end{equation}
In the absence of $m_{ph}$, Eq. (\ref{eq:av chi_gen1}) implies that $k \cdot q$ is simply related to the quantum parameter, namely  $\chi = \frac{ea (k \cdot q)}{m^3}$. Consequently, the familiar Volkov-Ritus expression is recovered
\begin{equation}
z = \frac{\xi^2 \sqrt{1+\xi^2}}{\chi} \sqrt{u \left( u^V_{s,max} - u \right)}
\label{eq:av z_vol}
\end{equation}
where (\ref{eq:av xi_def}, \ref{eq:av m_star}) were used as well.

\section{the continuum approximation}
As explicitly demonstrated in Sec. \rom{5}, the emission spectrum is composed of $s \rightarrow \infty$ harmonics. As a matter of fact, the number of harmonics with nonlegligible probability depends upon the non-linearity parameter $\xi$. 
In the Volkov-Ritus rate, for instance, the spectrum peak corresponds to $s \propto \xi^3$ in the strong field regime, $\xi \gg 1$.
If the emission spectrum consists of a large number of overlapping harmonics, it becomes a continuous function. 
In the following, we seek for the continuous limit of the rate obtained above (\ref{eq:av rate2}, \ref{eq:av fin_mat}).
For the sake of simplicity, the discussion is limited to the case of $q \cdot a_1  =0$, where the $\varphi$ dependence vanishes and the expression for $z$ is simpler (\ref{eq:av simple_z}). 

The essence of this approximation is the replacement of the sum over the harmonics by an integral, i.e.
\begin{equation}
\sum_s \rightarrow \int{ds}.
\end{equation}
As a result, (\ref{eq:av rate2}, \ref{eq:av fin_mat}) take the form (after the integration over $\varphi$)
\begin{equation}
\frac{dW}{du}=\frac{ m^2e^2}{16 \pi q_0 (1+u)^2} \int{ds} \eta_s F(s,u),
\label{eq:av fin_W}
\end{equation}
where
\begin{multline}
F \equiv -2 J_s^2 (z) \left( 2 + \frac{u-3}{2(u+1)} \frac{s^2 m_{ph}^2}{m^2} \right) \\ + 4 \xi^2 \left[ \left(\frac{s^2}{z^2}-1 \right) J_s^2(z)+ {J'}_s^2(z) \right].
\label{eq:av F_i}
\end{multline}
In addition, the relation between a Bessel function with large index ($s \gg 1$) to the Airy function \cite{watson} is employed
\begin{equation}
J_s(z) \approx  \left( \frac{2}{s}  \right)^{1/3} Ai(y),
\label{eq:av bessel_id}
\end{equation}
where $Ai(y) \equiv (1/\pi) \int_0^{\infty}{dt} \cos (t^3/3+yt)$ and
\begin{equation}
y \equiv \left( \frac{s}{2}  \right)^{2/3} \left( 1 - \frac{z^2}{s^2} \right).
\label{eq:av y_def}
\end{equation}
Let us write down the expressions appearing in (\ref{eq:av F_i}) in terms of $y$
\begin{equation}
\left(\frac{s^2}{z^2}-1 \right) J_s^2(z)=\left(\frac{2}{s} \right)^{4/3} \frac{ y}{1-(2/s)^{2/3} y} Ai^2(y)
\label{eq:av y_ident1}
\end{equation}
and
\begin{equation}
{J'}_s^2(z) = \left( \frac{2}{s} \right)^{4/3}Ai'^2(y) \sqrt{1 - y\left(\frac{2}{s} \right)^{2/3} }.
\label{eq:av y_ident2}
\end{equation}
Since the emission is attributed mainly to $z \rightarrow s$, we have $ y\left(\frac{2}{s} \right)^{2/3} \ll 1$ and therefore the root in (\ref{eq:av y_ident2}) and the denominator in (\ref{eq:av y_ident1}) may be approximated by 1. 
Finally, we obtain
\begin{multline}
F \approx - 4 \left( \frac{2}{s} \right)^{2/3} Ai^2 (y)  \\ +  4 \left( \frac{2}{s} \right)^{4/3} \xi^2 \left[ y Ai^2(y) + Ai'^2(y) \right].
\label{eq:av fin_F}
\end{multline}

The continuum approximation is applicable if the harmonics overlap to create a continuous spectrum. 
In order to formulate the continuum condition, a better understanding of a single harmonic structure is required.
The contribution of a single harmonic, as derived above, is a function of the corresponding Bessel function and its derivative $J_s(z), J'_s(z)$. The argument $z$ is a function of $u$, and one may prove that it lies in the range $0<z<z_{max}$. In the Volkov-Ritus case, the maximal value of $z$ is
\begin{equation}
z^V_{max} = s([1-1/(2\xi^2)], 
\label{eq:av z_max_v}
\end{equation}
as was derived in \cite{ritus}. For the new solution, (\ref{eq:av simple_z}, \ref{eq:av sin_u}, \ref{eq:av D_s}) imply that $z_{max}$ is given by 
\begin{equation}
z_{max} = \frac{ea p_{out}}{ k \cdot q+s m_{ph}^2} \frac{\eta_s \left(\eta_s \kappa_s +1 \right)}{\sqrt{\eta^4_s \kappa^2_s -1}}.
\end{equation}
One may show analytically that it always falls behind $z^V_{max}$. 

An illustration of the Bessel function for large $s$ appears in Fig. 1.
\begin{figure}[h]
  \begin{center}
   	   \includegraphics[width=0.5\textwidth]{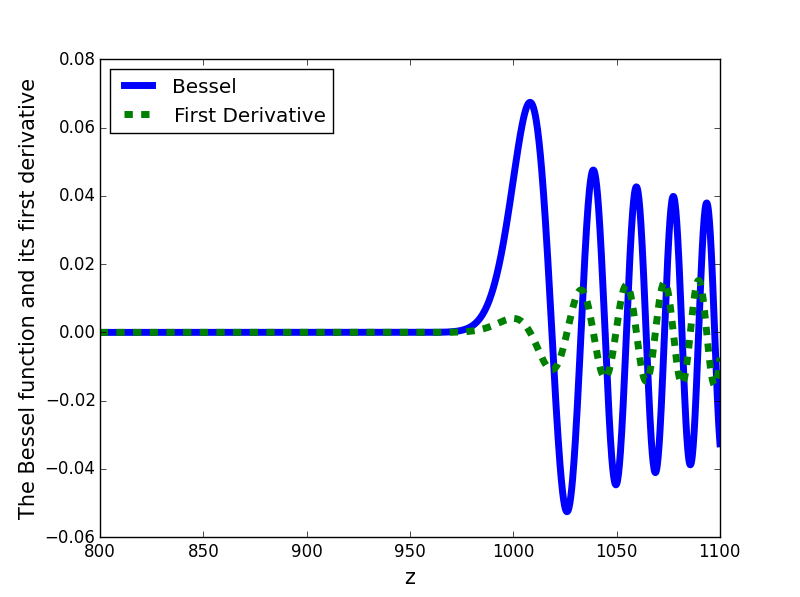}
      \caption{(color online). The Bessel function and its derivative ($J_s(z), J'_s(z)$) for $s=1000$.}
  \end{center}
\end{figure}
As can be seen, the Bessel function and its derivative vanish through most of the range $0<z<s$, but rise abruptly near $z=s$. Hence, the contribution to the emission comes from this region.
Let us estimate the argument $z_r$ for which the function starts rising.
For this purpose, the relation between the Bessel function and the Airy function (\ref{eq:av bessel_id}) is invoked once again.
The Airy function is approximately zero if its argument satisfies $y_r>3$. Accordingly, $z_r$ obeys 
\begin{equation}
y_{r} = 3 =  \left( \frac{s}{2}  \right)^{2/3} \left( 1 - \frac{z_{r}^2}{s^2} \right).
\label{eq:av cutoff1}
\end{equation}
For the Bessel function plotted above ($s=1000$), this estimation yields $z_r = 976$, in agreement with Fig. 1.
Since for large s we have $z_r \approx s$, a quantity measuring the distance bewtween $z_r$ and $s$ is introduced
\begin{equation}
\epsilon^r_{s} \equiv \frac{s - z_{r}}{s}.
\label{eq:av z_r}
\end{equation}
Substituting (\ref{eq:av z_r}) into (\ref{eq:av cutoff1}), $\epsilon^r_{s}$ is obtained
\begin{equation}
\epsilon^r_{s} = \frac{3}{2}   \left( \frac{2}{s}  \right)^{2/3}. 
\label{eq:av eps_r}
\end{equation}
The deviation of $z_{max}$ with respect to $s$ is denoted similarly by
\begin{equation}
\epsilon^{max}_{s} \equiv \frac{s - z_{max}}{s}.
\label{eq:av eps_max}
\end{equation}
Due to (\ref{eq:av z_max_v}), the deviation of $z^V_{max}$ from $s$ may be readily obtained
\begin{equation}
\epsilon^{V}_{s} \equiv \frac{s - z^V_{max}}{s}= \frac{1}{2 \xi^2}.
\label{eq:av eps_V}
\end{equation}

Having discussed the Bessel function behavior in the relevant regime, the width of a given harmonic may be readily obtained.
The $u$ values corresponding to the harmonic boundaries satisfy the equation $z(u)=z_r$. In terms of $z_{max}$, the argument $z$ given in (\ref{eq:av simple_z}) may be written as
\begin{equation}
z(u) = \frac{z_{max}}{\bar{D}_s} \sqrt{\bar{A}_s u^2+\bar{B}_s u+\bar{C}_s}.
\label{eq:av simple_z2}
\end{equation}
Employing (\ref{eq:av simple_z2}), the equation for the harmonic boundaries reads
\begin{equation}
\bar{A}_s u^2 + \bar{B}_s u +\bar{C}_s = \bar{D}^2_s \left(\frac{1-\epsilon^r_{s}}{1-\epsilon^{max}_{s}} \right)^2.
\label{eq:av ABCD}
\end{equation}
Notice that $z_{max}, z_r$ were replaced by $\epsilon^r_s, \epsilon^{max}_s$ according to Eqs. (\ref{eq:av eps_r}, \ref{eq:av eps_max}).
The difference $\Delta u_s$ between the solutions $u_{s,2},u_{s,1}$ of (\ref{eq:av ABCD}), corresponding the harmonic width, is
\begin{equation}
\Delta u_s = \frac{1}{\bar{A}_s} \sqrt{\bar{B}_s^2 - 4\bar{A}_s \left[ \bar{C}_s - \bar{D}^2_s \left(\frac{1-\epsilon^r_{s}}{1-\epsilon^{max}_{s}} \right)^2 \right]}.
\end{equation}
With the above expression at hand, the continuum condition may be quantitatively formulated.
As mentioned before, the spectrum is continuous if the harmonics overlap. It occurs if the spacing between two following harmonics is much smaller than the harmonic width, i.e.
\begin{equation}
\frac{\Delta u_s}{u^m_{s+1} - u^m_{s}} \gg 1.
\label{eq:av cont_cond}
\end{equation}
The inequality (\ref{eq:av cont_cond}) is harmonic dependent. In order to use the continuum formula, (\ref{eq:av cont_cond}) has to hold for all the harmonics with nonnegligible contribution to the spectrum.

Now let us show how, for vanishing $m_{ph}$, the expressions derived above (\ref{eq:av fin_W}, \ref{eq:av fin_F}) recover the Volkov-Ritus continuum approximation.
In the Volkov limit, the maximal value of the Bessel argument is (\ref{eq:av z_max_v}), corresponding to $y = \left[u/(2\chi) \right]^{2/3}$.
Expanding $y$ in the vicinity of this point, (\ref{eq:av y_def}) becomes (as was shown in \cite{ritus})
\begin{equation}
y = \left(\frac{u}{2\chi} \right)^{2/3} \left[ 1+ \tau^2 \right].
\label{eq:av y_tau}
\end{equation}
where the expansion parameter $\tau$ is related to $s$ by
\begin{equation}
\tau \equiv \xi \left( \frac{s \chi}{\xi^3 u} - 1-\frac{1}{2x^2} \right).
\label{eq:av y_tau}
\end{equation}
The substitution of (\ref{eq:av fin_F}, \ref{eq:av y_tau}) into (\ref{eq:av fin_W}) yields the Volkov-Ritus continuum approximation
\begin{multline}
\frac{dW}{du} = \frac{e^2m^2}{2 \pi^3 q_0} \int{d \tau} \frac{ \left(u/2 \chi \right)^{1/3}}{(1+u)^2} \bigl[ -Ai^2(y) + \\ \left(\frac{2u}{\chi} \right)^{2/3} \left[ yAi^2(y) + Ai'^2(y) \right] \bigr].
\label{eq:av CC}
\end{multline}
The expression above coincides with the rate of a particle in a constant crossed field. It should be mentioned that in this case, since the field frequency is assumed to go to zero, one may replace $q_0$ with $\Pi_0$ (see \cite{ritus}).

\section{the emission spectrum}
In the previous sections, the rate $W$ was calculated as a function of the invariant dimensionless variables $u, \varphi$. In practice, we are interested in the spectrum of the emitted power $P$ as a function of the outcoming photon energy $\omega'$. 
In the following, the transformation between these quantities is discussed. Since the numerical results appearing in this work were calculated for problems without $\varphi$ dependence (see the following section), it is omitted from this discussion as well.
The emitted power spectrum is given by
\begin{equation}
\frac{dP}{d\omega'} = \sum_s{} \omega'\frac{dW_s}{d \omega'}  = \sum_s{} \omega' \frac{dW_s}{du} \frac{du}{d\omega'}.
\end{equation}
The relation between $u$ and $\omega'$ stems from the definition (\ref{eq:av u_def1})
\begin{equation}
u = \frac{\omega'}{E^{lab}_s - \omega'} 
\label{eq:av omega_u}
\end{equation}
where $E^{lab}_s = q^{lab}_0+s\omega=\omega' + {q^{lab}_0}'$ is the total energy in the laboratory frame. The derivative is given by
\begin{equation}
\frac{du}{d \omega'} =\frac{1}{E^{lab}_s - \omega'} + \frac{\omega'}{\left(E^{lab}_s - \omega' \right)^2}  =  \frac{E^{lab}_s}{\left( E^{lab}_s- \omega' \right)^2}.
\label{eq:av omega_u}
\end{equation}
where the derivative of $E^{lab}_s$ with respect to $\omega'$ vanishes.
If $q^{lab}_0 \gg s\omega$ for all the harmonics with nonegligible contribution to the spectrum, which holds for our calculation, the energy $E^{lab}_s$ is the same for all the harmonics. As a result, $\omega' \frac{du}{d\omega'} $ may be extracted from the summation over $s$
\begin{equation}
\frac{dP}{d\omega'} = \frac{\omega'}{\left( q^{lab}_0 - \omega' \right)^2}  \frac{dW}{du}.
\label{eq:av p_omega}
\end{equation}
This is the final expression relating $dW / du$ obtained earlier to the actual measurable spectrum $dP/ d \omega'$.

To conclude the analytical part of the paper, let us summerize the final expressions obtained so far. The emission probabilty is given by (\ref{eq:av rate2}), the matrix element by (\ref{eq:av fin_mat}) and the variable $z$ is calculated using (\ref{eq:av z_def},\ref{eq:av alpha_2},\ref{eq:av alpha_1}). In the continuum approximation, these expressions are replaced by (\ref{eq:av fin_F}, \ref{eq:av fin_W}, \ref{eq:av y_def}, \ref{eq:av simple_z}). The relation between the emission probabilty and the spectrum in the laboratory frame was obtained in this section, Eq. (\ref{eq:av p_omega}).

\section{numerical results}
\subsection{Optical laser}
In the following, the radiation emitted by an electron interacting with a rotating electric field (in the lab frame) is numerically investigated. The field vector potential is given by (\ref{eq:av A_cl2}) and the wave vector is $k^{lab} = (m_{ph},0,0,0)$, see Sec. \rom{3} for details. It yields a homogenous electric rotating in the $x-y$ plane with frequency $\omega = m_{ph}$. As discussed in Sec. \rom{2}, it approximates the field in the vicinity of the antinode of a standing wave created by counterpropagating beams. The electron initial asymptotic momentum takes the form $p^{lab} = (p_0,0,0,p_z)$, and the corresponding quasi-momentum $q$ is related to the asymptotic momentum by (\ref{eq:av q_def}). The electron initial momentum was chosen to be perpendicular to the field plane in order to avoid dependence on $\varphi$ (since $q \cdot a_1= q \cdot a_2 =0$; see Eq. (\ref{eq:av simple_z})). The laser photons energy is $\omega = m_{ph} = 1.6eV$, corresponding to Ti-Sapphire. The intensity was chosen to be the present-days record \cite{yanovsky}, $I=10^{22}W/cm^2$, leading to a normalized amplitude of $\xi = 50$.

In Figs. 2, 4-7 the emitted photon spectrum is presented for several asymptotic momentum values. 
The solid line corresponds to the new solution and was calculated by (\ref{eq:av rate2},\ref{eq:av fin_mat}). It should be mentioned that for the parameters of Figs. 2,4,5 the continuum approximation (\ref{eq:av fin_W}, \ref{eq:av fin_F}, \ref{eq:av simple_z}) is adequate and gives exactly the same spectrum as the full calculation.
Our reference model (represented by the dashed line) is the Volkov-Ritus rate in the constant crossed field limit (\ref{eq:av CC}, \ref{eq:av z_vol}). As discussed in the introduction, this model is commonly assumed to be adequate for an arbitrary field configuration given that (\ref{eq:av assump}) is satisfied, which holds in the cases under consideration. The evaluation of (\ref{eq:av CC}) requires 3 quantities: $\chi,\xi$ and $\Pi_0$. 
The quantity $\Pi_0$ is taken from (\ref{eq:av Pi_f}) and $\chi$ is given in (\ref{eq:av chi_gen1}). Since the configuration considered here satisfies ${A^{cl}}' \cdot p =0$, the quantum parameter takes the simplified form
\begin{equation}
\chi  = \frac{ea(k \cdot q) }{m^3}.
\end{equation}
Notice that even though $\Pi_{\mu}$ is time dependent, $\chi$ is constant in time and changes only due to the emission process.

\begin{figure}[h]
  \begin{center}
   	   \includegraphics[width=0.5\textwidth]{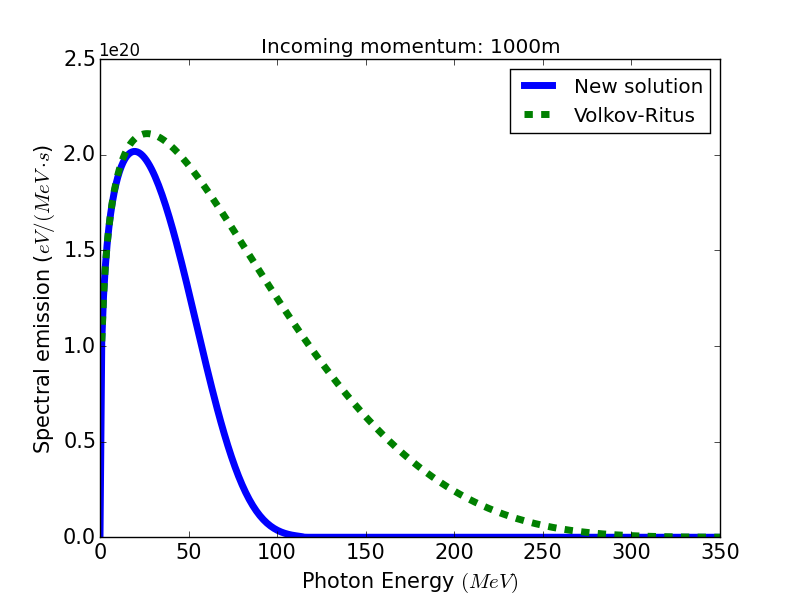}
      \caption{(color online). The spectral emission according to the new solution (solid curve) and the Volkov-Ritus rate (dashed curve), for $\xi=50, \omega=1.6eV, \textbf{p}^{lab} = (0,0,1000m)$.}
  \end{center}
\end{figure}

Fig. 2 shows the emitted spectrum for $p^{lab}_z = 1000m $. Namely, the electron is accelerated towards the laser with an energy of $ 0.5GeV$. Such conditions may be achieved either by a standard accelerator (such as SLAC \cite{bamber}) or by a laser-plasma accelerator \cite{esarey}. 
Since $\chi = 0.15$, this configuration lies in the quantum regime.
The two models coincide for soft outcoming photons ($<30MeV$) but for higher energies the new model decreases much faster. To be particular, the spectrum corresponding to the new solution dies out at $\omega'_* \approx 100MeV $ while for Volkov-Ritus we have ${\omega'}^V_* \approx 300MeV $. The $*$ symbol stands for the cutoff energy, meaning that the emission for $\omega' > \omega'_*$ is neglegible (roughly speaking, less than one percent of the spectrum peak value). Moreover, the total emitted power is $P = 1.1\cdot 10^{22}eV/s, P^V = 2.5 \cdot 10^{22}eV/s$ respectively, namely twice larger for the Volkov-Ritus calculation.

\begin{figure}[h]
  \begin{center}
   	   \includegraphics[width=0.5\textwidth]{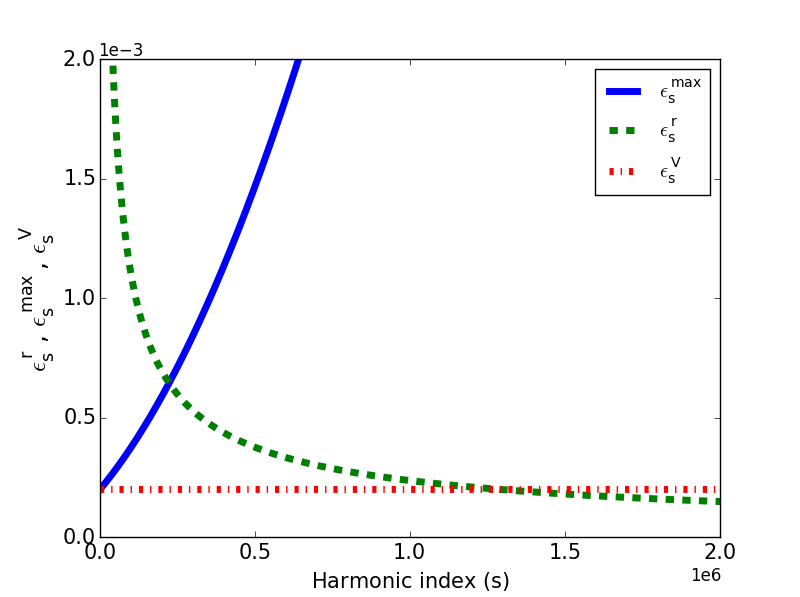}
      \caption{(color online) The dimensionless quantities $\epsilon^{max}_s, \epsilon^{r}_s$ and $\epsilon^{V}_s$ as a function of the harmonic index $s$. The solid line stands for $\epsilon^{max}_s$, representing the normalized deviation of the Bessel argument $z_{max}$ from $s$ and given in (\ref{eq:av eps_max}). The dashed line stands for $\epsilon^{r}_s$, representing the normalized deviation of the Bessel argument $z_{r}$ from $s$ and given in (\ref{eq:av eps_r}).
The dot-dashed line stands for $\epsilon^{V}_s$, representing the normalized deviation of the Bessel argument $z^V_{max}$ from $s$ and given in (\ref{eq:av eps_V}). }. 
  \end{center}
\end{figure}

In order to account for the spectral difference exhibited in Fig. 2, we seek a theoretical estimation for the cutoff harmonic $s_*$. 
In Sec. \rom{7} we have seen that the Bessel function of a given index $s$ goes to zero if its argument $z$ satisfies $z<z_r=s (1- \epsilon^r_{s})$. As a result, if the maximal argument of a given $s$, $z_{max} = s (1- \epsilon^{max}_{s})$, is smaller than $z_r$, the contribution of this harmonic should be negligible.
This insight enables us to write down the cut-off condition, $\epsilon^{max}_{s} = \epsilon^r_{s}$, satisfied by the cutoff harmonic $s_*$.
Fig. 3 shows the dependence of $\epsilon^r_{s}, \epsilon^{max}_{s}$ and $\epsilon^V_{s}$ on the harmonic index $s$ for the same physical parameters. According to the intersection points, one can deduce the cutoff harmonic: $s_* \approx 2.3 \cdot 10^5$ for the new model and $s^V_* \approx 1.3 \cdot 10^6$ for Volkov-Ritus. 

Now let us calculate the energy $\omega'_*$ corresponding to this harmonic and compare it with the one inspected from the numerical spectrum exhibited in Fig. 2. For this purpose, several quantites have to be evaluated:
$E_s = 62.3m, \kappa_s = 4.6, \eta_s = 1.001$, where Eqs. (\ref{eq:av Es2_def}, \ref{eq:av kappa_def}, \ref{eq:av alpha_fin}) were used.
It allows us to calculate the $u$ value for which this harmonic contributes (\ref{eq:av u_m}), namely $u^m_{s} \approx 0.27$. Plugging it into (\ref{eq:av omega_u}), one finds that the spectrum should die out at $\omega'_* \approx 110MeV$. An analogous procedure for the Volkov-Ritus model yields $\kappa_s = 1.67, u^m_s = 1.5$ and thus ${\omega'}^V_* \approx 305MeV$. Both estimation are in excellent agreement with the numerical calculation of Fig. 2. 

According to the numerical calculation, this effect (i.e. a lower cutoff for the new model spectrum) decreases for increasing asymptotic momentum. 
The spectra of the two models coincide for $p_z > 1.5 \cdot 10^4m$. 
In the following the opposite limit is explored - the asymptotic momentum is gradually reduced to non-relativistic values.

\begin{figure}[h]
  \begin{center}
   	   \includegraphics[width=0.5\textwidth]{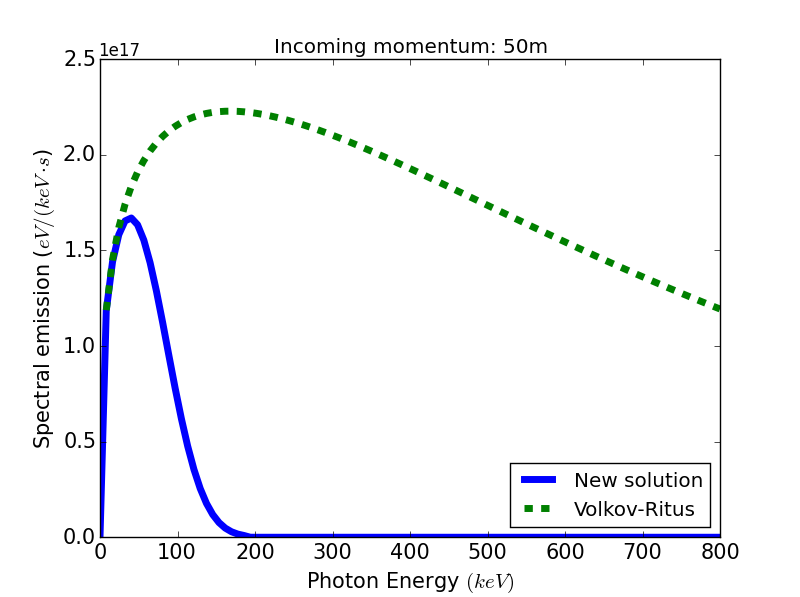}
      \caption{(color online). The spectral emission according to the new solution (solid curve) and the Volkov-Ritus rate (dashed curve), for $\xi=50, \omega=1.6eV, \textbf{p}^{lab} = (0,0,50m)$.}
  \end{center}
\end{figure}
\begin{figure}[h]
  \begin{center}
   	   \includegraphics[width=0.5\textwidth]{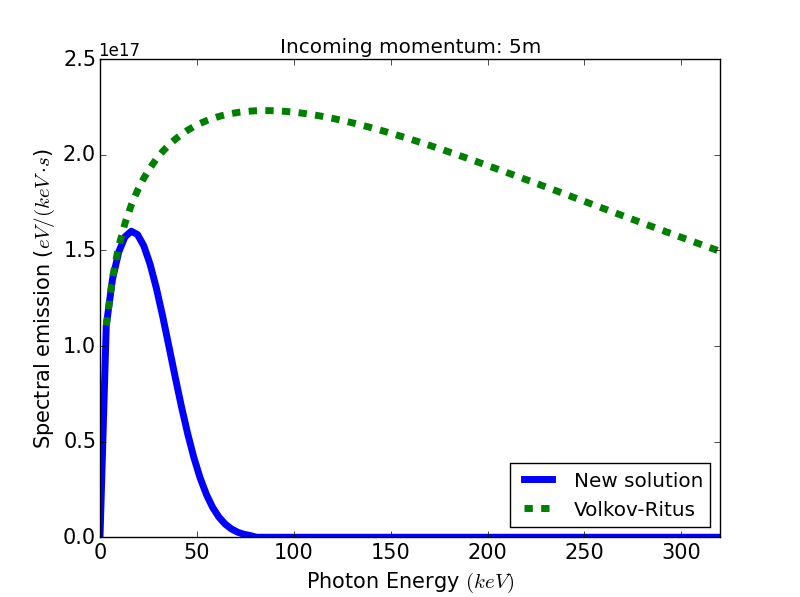}
      \caption{(color online).  The spectral emission according to the new solution (solid curve) and the Volkov-Ritus rate (dashed curve), for $\xi=50, \omega=1.6eV, \textbf{p}^{lab} = (0,0,5m)$.}
  \end{center}
\end{figure}

In Fig. 4 the momentum is equal to the field amplitude, i.e. $p^{lab}_z = 50m$. This case is of special interest as it is the maximal momentum possible if the particle is accelerated by the field itself (without using an external accelerator). The difference between the models is analogous to the one appearing in Fig. 2 but more pronounce - now the total emitted power is $P = 0.15\cdot 10^{20} eV/s$, $P^V = 2.3 \cdot 10^{20} eV/s$ and the cutoff is $\omega'_* = 200keV$, ${\omega'}^V_* = 4000keV $ respectively. Notice that due to the lower asymptotic momentum, the spectral cutoff and the emitted power of both models are reduced in orders of magnitude compared to those in Fig. 2.

In Fig. 5 the incoming momentum is an order of magnitude smaller than the field amplitude ($p^{lab}_z = 5m$). 
the total emitted power is $P = 0.06 \cdot 10^{20} eV/s$, $P^V = 1.17 \cdot 10^{20} eV/s$ and the cutoff is $\omega'_* = 75keV$, ${\omega'}^V_* = 2000keV $ respectively. That is to say, the emitted power corresponding to the new solution is lower by a factor of 20 than the power predicted by the Volkov-Ritus model.
It can be seen that as compared to Fig. 4, the spectral shape as well as the maximum value of $dP/d \omega'$ remain the same for both models. The difference is in the lower cut-off and, as a result, the total emitted power. The reason lies in the lower value of $k \cdot q$ and therefore of $\chi$.

\begin{figure}[h]
  \begin{center}
   	   \includegraphics[width=0.5\textwidth]{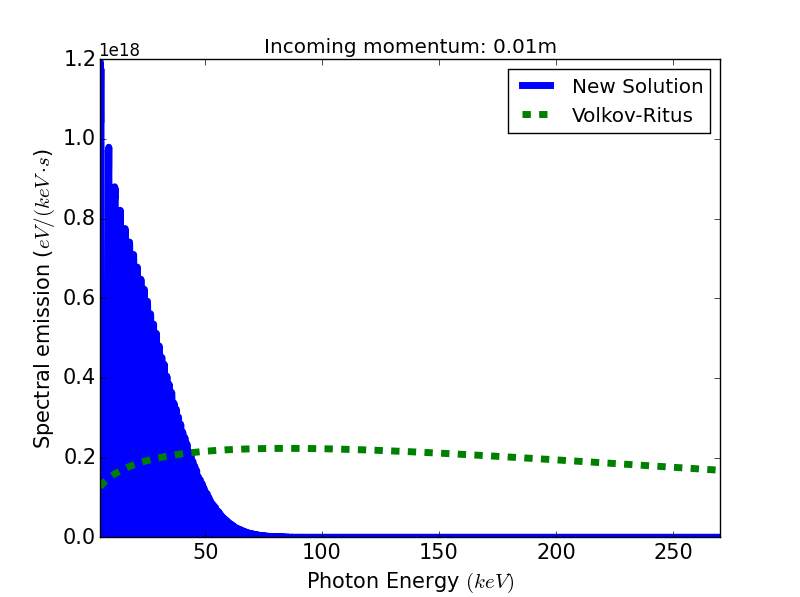}
      \caption{(color online). The spectral emission according to the new solution (solid curve) and the Volkov-Ritus rate (dashed curve), for $\xi=50, \omega=1.6eV, \textbf{p}^{lab} = (0,0,0.01m)$.}
  \end{center}
\end{figure}

In Fig. 6 the incoming momentum is decreased even lower ($p^{lab}_z=0.01m$), giving rise to a novel phenomenon. The harmonics width becomes smaller than the spacing between following harmonics, and consequently the spectrum is no longer continuous but takes a comb-like structure. It corresponds to the breakdown of the continuum condition derived in Eq. (\ref{eq:av cont_cond}).

In the following we suggest two qualitative explanations for this phenomenon. The classical one is that for negligible $p_z$ values, the electron motion follows the vector potential, as can be inferred from (\ref{eq:av Pi_f}). Consequently, the motion is circular. In this case, as was found long ago by Schott, the particle radiates discrete harmonics \cite{schott,landau2}. It may also be readily seen from the energy-momentum conservation (\ref{eq:av delta_cons}). In the classical case $q \cdot q  \approx q \cdot q'$. Therefore,
\begin{equation}
s q \cdot k = q \cdot k'.
\end{equation}
Writing the above equation in the laboratory frame, one obtains
\begin{equation}
\omega'= \frac{s \omega q^{lab}_0}{q^{lab}_0 - |\textbf{q}|^{lab} \cos \gamma} \approx s \omega.
\end{equation}
where $\gamma$ is the angle between $\textbf{q}'$ and $\textbf{k}'$ in the laboratory frame. Since the second term in the denominator is negligible with respect to the first, the emitted harmonics are simply multipilcation of the original one.

From the quantum point of view, we have seen in Sec. \rom{4} that in the center of mass frame the angle of the outcoming photon may get any value but its energy has a certain value $k_0'$. Transforming to the lab frame of reference, different emission angles correspond to different Lorentz transformations, giving rise to an energy spread. As a result, the closer is the laboratory frame to the center of mass frame of a given harmonic, the narrower is its width.
This condition is achieved by lowering the incoming momentum $p_z$.

Fig. 7,8 are zoom-in presentations of Fig. 6 in different spectral regions. Fig. 7 shows a range of $8eV$ in the soft part of the spectrum, and Fig. 8 shows the same range near the peak. They demonstrate that the harmonics width is extremely small for low energy photons and increases with the photon energy, in accordance with the explanation above.
\begin{figure}[h]
  \begin{center}
   	   \includegraphics[width=0.5\textwidth]{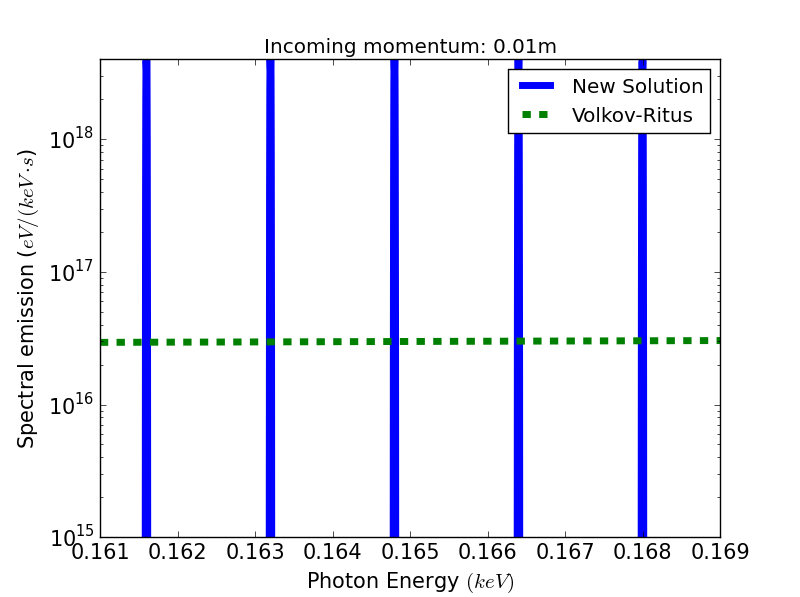}
      \caption{(color online). Zoom in of Fig. 6 in the range 161-169eV.}
  \end{center}
\end{figure}
\begin{figure}[h]
  \begin{center}
   	   \includegraphics[width=0.5\textwidth]{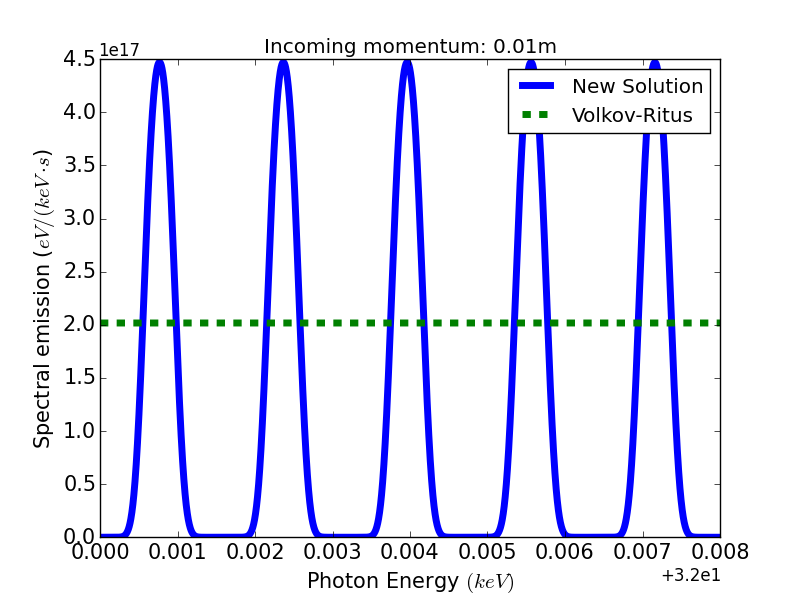}
      \caption{(color online). Zoom in of Fig. 6 in the range 32-32.008keV.}
  \end{center}
\end{figure}

\subsection{X-ray laser}
In this subsection another possible experimental set up is discussed. The optical laser is replaced by an X ray laser with the parameters 
of the LCLS X-ray Free Electron Laser facility \cite{LCLS}, i.e. $I=4 \cdot 10^{20}W/cm^2$ and $\omega = m_{ph} = 10keV$ corresponding to $\xi = 2 \cdot 10^{-3}$.
Due to the small value of $\xi$, the emission involve only the two first harmonics and the continuum approximation could not be used. 
As a result, the constant crossed field condition (\ref{eq:av assump}) is not satisfied and the Volkov-Ritus model is inapplicable in a rotating electric field even according to the nikishov-Ritus assumption described in the introduction. Nevertheless, in the absence of any other adequate model, it is used as a reference for our prediction.
Consequently, the Volkov-Ritus rate was calculated employing the full expression (\ref{eq:av fin_W}, \ref{eq:av fin_Vol}, \ref{eq:av z_vol}) instead of the continuum expression used for the optical laser above.

\begin{figure}[h]
  \begin{center}
   	   \includegraphics[width=0.5\textwidth]{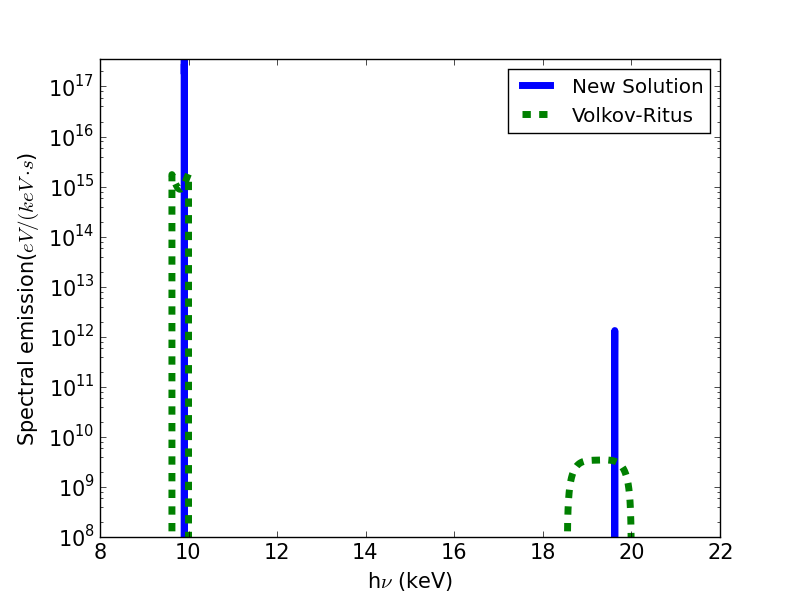}
      \caption{(color online). The spectral emission according to the new solution (solid curve) and the Volkov-Ritus rate (dashed curve), for $\xi=2 \cdot 10^{-3}, \omega=10keV, \textbf{p}^{lab} = (0,0,10^{-4}m)$.}
  \end{center}
\end{figure}

Fig. 9 compares the emission predicted by the new solution to the Volkov-Ritus for non-relativistic asymptoric momentum $p^{lab}_z=10^{-4}m$.
According to (\ref{eq:av alpha_s2}, \ref{eq:av kappa_def}, \ref{eq:av u_m}), the first and second harmonics correspond to $u^m_s = 0.01, 0.02$ respectively. Hence, the scattering is in the weakly quantum regime.
The energy contained under the curves is roughly identical, as opposed to the optical laser calculations above. The difference in this case lies in the width of the harmonics. The mechanism behind the narrowing is the same as encountered in Fig. 6-8. The second harmonic is of special interest for several reasons. First, the width differences are more pronounce for this harmonic. Second, it stems from the nonlinearity of the interaction and could be seen for strong fields only. Third, a measurement of the second harmonic under similar conditions was recently demonstrated \cite{nature_compton}. It should be stressed that our result could not be compared with this specific experiment since it was not carried out in a standing wave. However, it shows that an exprimental test of our calculation is fissible with nowadays facilities.
The advange of this experimental sut-up over the former one (the optical laser) is that the quantum regime can be achieved without an external accelerator.

\section{conclusions}
In this article, the non-linear Compton rate in a rotating field was investigated for the first time. For the sake of this purpose, we employed a novel analytical solution to the Klein-Gordon equation describing a particle in the presence of this field configuration, lately derived by the authors \cite{myPaper3}. Closed analytical expression for the relevant matrix element was obtained in Eq. (\ref{eq:av fin_mat}). 

Furthermore, we have shown that for strong fields ($\xi \gg 1$) and initial asymptotic momentum satisfying the condition (\ref{eq:av cont_cond}), the spectrum may be approximated by a continuous function instead of discrete harmonics sum. This is a generalization of the familiar continuum approximation of the Volkov-Ritus rate \cite{nikishov}. The final expression (\ref{eq:av fin_W}, \ref{eq:av fin_F}) is easy to calculate and may be employed in kinetic laser-plasma calculations.

Numerical calculations of the emitted photon spectrum according to the new rate were carried out and compared with the Volkov-Ritus rate.  Physical parameters corresponding to the state-of-the-art facilities of both optical lasers ($I=10^{22}W/cm^2, \omega = 1.6eV$) and XFEL lasers ($I=4 \cdot 10^{20}W/cm^2, \omega = 10keV$) were chosen.

In the first case (optical laser), the Volkov-Ritus rate reduces to the constant crossed field rate, frequently used in QED-PIC simulations. The following points arise from the comparison:
\begin{itemize}
\item
The deviation between our expression and the Volkov-Ritus one in the total emitted power grows for decreasing incoming particle asymptotic momentum $p^{lab}_z$ and amounts to a factor of 20. In addition, the spectrum cutoff energy is considerably lower for our new solution. As an explanation for this phenomenon, a semi-analytical way to determine the cut-off energy is suggested and achieves a good agreement with the calculated spectrum.
The discrepancy between our model and the Ritus-Volkov one decreases for increasing $p^{lab}_z$. The value above which both models coincide was found by numerical means.
\item
For $p^{lab}_z \ll ea$ the energetic width of the harmonics that compose the spectrum falls beneath the spacing between them. As a result, the spectrum structure becomes comb-like, as opposed to the continuous shape of the Volkov-Ritus rate under these conditions. An intuitive explanation for the discrepancy is suggested. This phenomenon is a clear evidence to the imprint of the rotating frequency on the emission spectrum, as opposed to the constant field paradigm.
\end{itemize}

In the second case (X-ray laser), the emitted power according to the models under consideration was approximately equivalent. However, the new solution predicted much narrower harmonics. The mechanism behind this narrowing is the same as for the optical laser.
The importance of the X-ray laser set-up stems from the fact that it does not require high asymptotic momentum. As a result, it enables
experimental verification of our model in the quantum regime without using an accelerator (as opposed to the optical laser set-up).

To conclude, the above calculations may be exprimentally tested for present days laser systems. In addition they may be of great importance in the context of PIC-QED calculations of the QED cascade mechanism assumed to be measurable for the next generation laser facillities.

\appendix
\numberwithin{equation}{section}

\section{}
In the following, the relation (\ref{eq:av cos_u}) between $u$ and $ \cos\theta$ is explicitly derived.
Substituting $k'$ from the energy - mometum conservation (\ref{eq:av delta_cons}) into the definition of $u$ (\ref{eq:av u_def1}) we arrive at
\begin{equation}
u = \frac{s m^2_{ph} + k \cdot q}{k \cdot q'} -1.
\label{eq:av u1}
\end{equation}
Let us write down the explicit expression for $u$ in the center of mass frame. Using (\ref{eq:av k_init}, \ref{eq:av q_init}), the nominator reads
\begin{equation}
k \cdot q + sm^2_{ph}= k_0 q_0 + p_{in}^2 / s + sm^2_{ph}.
\end{equation}
Since $k_0^2 = (p_{in}/s)^2+m_{ph}^2$, the former equation becomes
\begin{equation}
k \cdot q + sm^2_{ph} =  k_0q_0 +sk_0^2.
\end{equation}
Employing the relation $E_s = sk_0 +q_0$, we find
\begin{equation}
k \cdot q + sm^2_{ph} =  k_0 E_s.
\label{eq:av kq_plus}
\end{equation}
We substitute (\ref{eq:av kq_plus}) into (\ref{eq:av u1}) and evaluate $k \cdot q'$ using (\ref{eq:av q_tag1}).
\begin{equation}
u = \frac{sk_0 E_s}{sk_0 q_0' - p_{in} p_{out}\cos \theta} - 1.
\end{equation}
In terms of $\eta_s$ (defined in (\ref{eq:av alpha_s1})) it becomes
\begin{equation}
u = \frac{ \eta_s E_s}{ \eta_s q_0' -  p_{out}\cos \theta} - 1.
\end{equation}
Hence, $\cos \theta$ can be obtained in term of $u$
\begin{equation}
\cos \theta = \frac{\eta_s}{ p_{out}} \left(  q_0'  -  \frac{E_s}{1+u}  \right).
\label{eq:av cos_u_pre}
\end{equation}
Substituting the expression (\ref{eq:av p_out}) for $p_{out}$ and using the follwing identity
\begin{equation}
q_0'  = \sqrt{p_{out}^2 + m_*^2}=\frac{E_s^2 + m_*^2}{2E_s}
\end{equation}
one finds
\begin{equation}
\cos \theta = \eta_s \left( \kappa_s - \frac{\kappa_s+1}{1+u} \right)
\end{equation}
where $\kappa_s$ was defined in (\ref{eq:av kappa_def}).

\section{}
This appendix is dedicated to the derivation of the matrix element expression (\ref{eq:av fin_mat}).
Substituting (\ref{eq:av M_th}) into (\ref{eq:av polar_sum}) we have
\begin{multline}
\frac{1}{2}\sum_{\epsilon'} |\mathcal{M}|^2 = -2e^2B^2 \left( m_*^2 + q \cdot q' \right) \\  +2Bk \cdot \left(q+q' \right) \left[\bar{\alpha}_1 \Re{(B_1)}  + \bar{\alpha}_2 \Re{(B_2)} \right] \\ -4eB \left(q +q' \right) \left[ a_1\Re{(B_1)}  + a_2 \Re{(B_2)} \right] \\ -  4e^4a^2 \left( |B_1|^2+|B_2|^2 \right)  + e | \bar{\alpha}_1 B_1 + \bar{\alpha}_2 B_2 |^2 m_{ph}^2 
\label{eq:av mat_el0}
\end{multline}
where $\Re$ denotes real part and the following definition is used
\begin{equation}
\bar{\alpha}_i \equiv \frac{e (a_i \cdot q)}{ (k \cdot q)}+\frac{e (a_i \cdot q')}{ (k \cdot q')}, \quad i=1,2.
\label{eq:av alpha_bar_def}
\end{equation}
It should be mentioned that since $\mathcal{M}_{\mu} \propto e^{is\phi_0}$, this constant exponent does not contribute to $\mathcal{M}^2$ and thus may be omitted. As a result, $B$ is real (while $B_1,B_2$ remain complex).

Before starting, let us examine the last term in Eq. (\ref{eq:av mat_el0}). It is of the same order of magnitude as the quantity $\delta$ assumed to be small in the new wavefunction derivation (see section \rom{3}), and thus may be neglected.
In order to simplify (\ref{eq:av mat_el0}), the term $k \cdot \left(q+q' \right) \bar{\alpha}_1$ is further worked out.
\begin{multline}
k \cdot \left(q+q' \right) \bar{\alpha}_1  =   e(a_1 \cdot q) + e(a_1 \cdot q') \\+e\left( a_1 \cdot q \right) \frac{k \cdot q'}{k \cdot  q} +e\left( a_1 \cdot q' \right) \frac{k \cdot q}{k \cdot  q'}
\label{eq:av eval1}
\end{multline}
where the explicit formula (\ref{eq:av alpha_bar_def}) for $\bar{\alpha}_1$ was used.
Due to the following alegbraic identities
\begin{equation}
\frac{k \cdot q}{k \cdot  q'} = 1 + \frac{k \cdot \left(q-q' \right) }{k \cdot q'},
\end{equation}
\begin{equation}
\frac{k \cdot q'}{k \cdot  q} = 1 - \frac{k \cdot \left(q-q' \right) }{k \cdot q},
\end{equation}
Eq. (\ref{eq:av eval1}) becomes 
\begin{multline}
k \cdot \left(q+q' \right) \bar{\alpha}_1  =   2e(a_1 \cdot q ) \\+ 2e(a_1 \cdot q') -\alpha_1 \left(q - q' \right),
\label{eq:av identity10}
\end{multline}
where $\alpha_1$ was defined in Eq. (\ref{eq:av alpha_def}).
Analogously, we have
\begin{multline}
k \cdot \left(q+q' \right) \bar{\alpha}_2  =   2e(a_2 \cdot q ) \\ + 2e(a_2 \cdot q') -\alpha_2 \left(q - q' \right).
\label{eq:av identity11}
\end{multline}
Substituting (\ref{eq:av identity10}, \ref{eq:av identity11}) into (\ref{eq:av mat_el0}), several terms cancel out and we are left with
\begin{multline}
\frac{1}{2}\sum_{\epsilon'} |\mathcal{M}|^2 = -2e^2B^2 \left( m_*^2 + q \cdot q' \right) \\-  4e^4a^2 \left( |B_1|^2+|B_2|^2 \right)   \\+2eB (q-q') \left[ \Re{(B_1)}  \alpha_1+  \Re{(B_2)}  \alpha_2 \right].  
\label{eq:av mat_el1}
\end{multline}
In virtue of the identity \cite{ritus}
\begin{equation}
\alpha_1 B_1 + \alpha_2 B_2 = sB,
\end{equation}
one obtains
\begin{multline}
\frac{1}{2}\sum_{\epsilon'} |\mathcal{M}|^2 = -2e^2B^2 \left( m_*^2 + q \cdot q' \right) \\-  4e^4a^2 \left( |B_1|^2+|B_2|^2 \right)   +2e^2sB^2 k \cdot (q-q').
\end{multline}
Employing (\ref{eq:av kdq_tag}) we have
\begin{equation}
k \cdot (q-q') =\frac{u k \cdot q +sm_{ph}^2}{u+1}.
\end{equation}
As a result, the matrix element becomes
\begin{multline}
\frac{1}{2}\sum_{\epsilon'} |\mathcal{M}|^2 = -2e^2B^2 \left( m_*^2 + q \cdot q' \right) \\-  4e^4a^2 \left( |B_1|^2+|B_2|^2 \right)   \\+2e^2sB^2 \frac{u k \cdot q +sm_{ph}^2}{u+1}.
\end{multline}
In order to procede, let us express $q \cdot q'$ in terms of $u$. Multiplying (\ref{eq:av delta_cons}) by $q'$, one obtains
\begin{equation}
q \cdot q' = m_*^2+k' \cdot q' - s k \cdot q'.
\label{eq:av qq_tag}
\end{equation}
The center of mass energy may be written in terms of the outcoming 4-momenta
\begin{equation}
E_s^2 = (k' + q')^2 = m_*^2+2k' \cdot q'.
\label{eq:av Es2}
\end{equation}
Therefore, $k' \cdot q'$ reads
\begin{equation}
k' \cdot q'  = \frac{E_s^2 - m_*^2}{2}.
\label{eq:av k_tag_q_tag}
\end{equation}
The substitution of (\ref{eq:av k_tag_q_tag}, \ref{eq:av kdq_tag}) into (\ref{eq:av qq_tag}) yields
\begin{equation}
q \cdot q' = \frac{E_s^2 + m_*^2}{2}-  \frac{s^2 m^2_{ph} + s(k \cdot q)}{u+1}.
\label{eq:av qq_tag2}
\end{equation}
With the aid of (\ref{eq:av qq_tag2}), one gets
\begin{multline}
\frac{1}{2}\sum_{\epsilon'} |\mathcal{M}|^2 = -2e^2B^2 \left( 2m_*^2 + \frac{u-3}{2(u+1)} s^2 m_{ph}^2 \right) \\-  4e^4a^2 \left( |B_1|^2+|B_2|^2 \right).   
\label{eq:av mat_sum}
\end{multline}
Plugging $B,B_1,B_2$, defined in (\ref{eq:av B0_def}-\ref{eq:av B2_def}) into (\ref{eq:av mat_sum}) one obtains the final expression
\begin{multline}
\frac{1}{2}\sum_{\epsilon'} |\mathcal{M}_{s,\epsilon'}|^2 = -2e^2 J_s^2 (z) \left( 2m_*^2 + \frac{u-3}{2(u+1)} s^2 m_{ph}^2 \right) \\ + 4 e^4a^2 \left( \frac{s^2}{z^2} J_s^2(z)+ {J'}_s^2(z) \right).
\end{multline}

\section{}
In this appendix, an expression for the component $a_1$ of the vector potential (defined in (\ref{eq:av a12_def})) in the c.m.s frame is obtained. The most general expression is
\begin{equation}
a_1 = a(R_0,R_1,0,R_3).
\label{eq:av a1_gen1}
\end{equation}
For the moment, $R_0,R_1,R_3$ are unknown. Furtunately, $a_1$ is known to obey several identities, and thus relations between these coefficients may be deduced.
Since $a_1 \cdot k = 0$ we have
\begin{equation}
R_3 = R_0 \frac{\omega}{k_z} = R_0 \eta_s.
\end{equation}
Due to $a_1 \cdot a_1 = -a^2$ we have
\begin{equation}
R_1 = \sqrt{1 - R_0^2 \left( \eta_s^2 - 1 \right)}.
\end{equation}
Hence, (\ref{eq:av a1_gen1}) takes the form
\begin{equation}
a_1 = a\left(R_0,\sqrt{1 - R_0^2 \left( \eta_s^2 - 1 \right)},0,R_0 \eta_s \right).
\label{eq:av a1_gen2}
\end{equation}
The final step is to obtain $R_0$ in terms of the known quantity $q \cdot a_1$. For this purpose, we write $q \cdot a_1$ in the c.m.s frame
\begin{equation}
q \cdot a_1 = a R_0 \left( q_0 + s \omega   \right) = a R_0  E_s.
\end{equation}
Therefore
\begin{equation}
R_0  =  \frac{q \cdot a_1} {aE_s}.
\end{equation}

\end{document}